\documentclass[iop,apj]{emulateapj}

\usepackage{txfonts}
\usepackage{graphicx,bm}
\usepackage[colorlinks,citecolor=blue,urlcolor=magenta]{hyperref}
\usepackage{aas_macros}

\begin{document}

\shorttitle{The $\nu$ process}
\shortauthors{Sieverding et al.}

\title{The $\nu$ process in the light of an improved understanding of
  supernova neutrino spectra}

\author{A. Sieverding\altaffilmark{1,2},
  G. Mart\'inez-Pinedo\altaffilmark{2,1}, L. Huther\altaffilmark{1},
  K.~Langanke\altaffilmark{2,1}, and A. Heger\altaffilmark{3,4,5}}
\affiliation{\altaffilmark{1}Institut f{\"u}r Kernphysik
  (Theoriezentrum), Technische Universit{\"a}t Darmstadt,
  Schlossgartenstra{\ss}e 2, 64289 Darmstadt, Germany}
\affiliation{\altaffilmark{2}GSI Helmholtzzentrum f{\"u}r
  Schwerioneneforschung, Planckstra{\ss}e 1, 64291 Darmstadt, Germany}
\affiliation{\altaffilmark{3}Monash Centre for Astrophysics, School of
  Physics and Astronomy, Monash University, Victoria 3800, Australia}
\affiliation{\altaffilmark{4}Tsung-Dao Lee Institute, Shanghai 200240, China}
\affiliation{\altaffilmark{5}The Joint Institute for Nuclear
  Astrophysics, Michigan State University, East Lansing, Michigan
  48824, USA }

\begin{abstract}
  We study the neutrino-induced production of nuclides in explosive
  supernova nucleosynthesis for progenitor stars with solar
  metallicity including neutrino nucleus reactions for all nuclei with
  charge numbers $Z < 76$ with average neutrino energies in agreement
  with modern Supernova simulations.  Considering progenitors with
  initial main sequence masses between 13~M$_\odot$ and 30~M$_\odot$,
  we find a significant production of $^{11}$B, $^{138}$La, and
  $^{180}$Ta by neutrino nucleosynthesis, despite the significantly
  reduced neutrino energies.  The production of $^{19}$F turns out to
  be more sensitive to the progenitor mass and structure than to the
  $\nu$ process. With our complete set of cross sections we have
  identified effects of the $\nu$~process on several stable nuclei
  including $^{33}$S, $^{40}$Ar, $^{41}$K, $^{59}$Co, and $^{113}$In
  at the 10\% level. Neutrino-induced reactions contribute to a similar extent to the
  production of radioactive ${}^{26}$Al and increase the yield of $^{22}$Na by 50\%. Future
  $\gamma$~ray astronomy missions may reach the precision at which
  the contribution from the $\nu$~process becomes relevant. We find
  that the production of $^{22}$Na by the $\nu$~process could explain
  the Ne-E(L) component of meteoritic graphite grains. The $\nu$~process
  enhances the yield of $^{36}$Cl and we point out that 
  the resulting $^{36}$Cl/$^{35}$Cl ratio is in agreement with the 
  values infrerred for the early solar system. Our extended
  set of neutrino-nucleus interactions also allows us to exclude any
  further effects of the $\nu$ process on stable nuclei and to
  quantify the effects on numerous, hitherto unconsidered radioactive
  nuclei, e.g., $^{36}$Cl, $^{72}$As, $^{84}$Rb, and $^{88}$Y.
\end{abstract} 

\maketitle
\section{Introduction}
\label{sec:intro}

Astrophysical objects like stars, novae, or supernovae are the origin
of most of the elements in the
Universe~\citep{Burbidge.Burbidge.ea:1957,Cameron:1957}. Whereas the
likely nucleosynthesis processes associated with these objects have
been identified and a general understanding has been developed, many
details of their operation are still
unresolved~\citep{Wiescher.Kaeppeler.Langanke:2012,Thielemann.Arcones.ea:2011,Arnould.Takahashi:1999}.
This is due to limited computational capabilities to simulate
astrophysical objects and the full range of their varieties and to the
fact that the properties of the nuclides involved in the
nucleosynthesis processes are not known experimentally and have to be
modeled~\citep{Grawe.Langanke.Martinez-Pinedo:2007,Langanke.Schatz:2013}.

An important test for models that aim to describe how the elements
are formed is the reproduction the chemical composition of well-known
objects like the Solar System. The composition of the Solar System has
been determined with high accuracy
\citep{Lodders:2003,Asplund.Grevesse.ea:2009} but we still lack the
full understanding of the origin of all the isotopes that can be
observed since it likely involves contributions from a multitude of
events that are difficult to disentangle.  Primitive meteorites are
another source of information on the nucleosynthesis that has
contributed to the composition of our solar system. Isotopic anomalies
have allowed to prescribe pre-solar grains that reflect the
composition of the early solar system (ESS) to different astrophysical
scenarios. The presence of $^{44}$Ti and an excess of Si as well as
the isotopic ratios of Fe and Ni in SiC-X grains and graphite grains
have led to the conclusion that these grains originate from a
core-collapse supernova \citep{Amari.Hoppe.ea:1992}.  Detections of
gamma-rays from radioactive nuclei by space bound observatories like
INTEGRAL~\citep{Winkler.Diehl.ea:2011} are another invaluable tool to
determine sites of active nucleosynthesis and thereby advance our
understanding of astrophysical nucleosynthesis.  Such detection allows
for a snapshot view of the ongoing nucleosynthesis in our galaxy and,
provided a suited nuclear half-life, to relate the origin of the
nuclide to a specific astrophysical
source~\citep{Diehl.Halloin.ea:2006}. In cases where the observation
can be assigned to a particular supernova remnant, one can learn about
asymmetries in the explosion~\citep{Grefenstette.Harrison.ea:2014,Wongwathanarat.Janka.ea:2017}.
The prime nuclide for gamma-ray astronomy in recent years has been
${}^{26}\!$Al~\citep{Diehl:2013}. Its production has been associated with
several astrophysical sources~\citep[see][and references
therein]{Woosley.Hartmann.ea:1990}, however, in recent years evidence
has been brought
forward~\citep{Timmes.Woosley.ea:1995,Diehl.Timmes:1998,Diehl:2013}
that massive stars can account for most of the ${}^{26}\!$Al in the
galaxy.  Other gamma-ray astronomy candidates such as $^{22}$Na,
$^{44}$Ti, and $^{60}$Fe are also related to core-collapse
supernovae~\citep{Iyudin.Diehl.ea:1994,Timmes.Woosley.ea:1995,Woosley.Heger.Weaver:2002,Rauscher.Heger.ea:2002,Limongi.Chieffi:2006}.

It has long been recognized that neutrino-nucleus reactions are
essential for the synthesis of selected nuclides including $^{7}$Li,
$^{11}$B, $^{15}$N, $^{19}$F, $^{138}$La, and
$^{180}$Ta~\citep{Woosley.Hartmann.ea:1990,Heger.Kolbe.ea:2005} or can
contribute to the production of long-lived radioactive
nuclides~\citep{Woosley.Hartmann.ea:1990,Timmes.Woosley.ea:1995,Woosley.Heger.Weaver:2002,Rauscher.Heger.ea:2002}.
This is denoted as $\nu$~process and involves neutrinos of all
flavors, which emitted from the hot proto-neutron star (PNS) formed
after a supernova explosion, interact with nuclei as they pass through
the surrounding stellar matter.  At the same time, these outer layers
are heated up and compressed by the explosion shock-wave propagating
outward from the PNS and causing the ejection of the material.
Neutral-current reactions excite the nucleus to states above particle
thresholds so that the subsequent decay is accompanied by emission of
light particles (proton, neutron, or $\alpha$ particle). Due to the
relatively low energies of the neutrinos, charged-current reactions
can only be induced by electron-type neutrinos. This process can be
accompanied by light-particle emission if the $(\nu_\mathrm{e},\mathrm{e}^-)$ or
$(\bar{\nu}_\mathrm{e}, \mathrm{e}^+)$ reactions excite the daughter nucleus to levels
above particle thresholds. Hence there are two possible ways for the 
$\nu$~process to contribute to nucleosynthesis.  Firstly, rare nuclei, e.g., 
$^{11}$B, $^{19}$F, $^{138}$La, and $^{180}$Ta, can be produced directly as
daughter products of neutrino-induced reactions on abundant nuclei.  And secondly,
neutrino spallation reactions increase the amount of light particles required
to synthesize some nuclides such as $^{7}$Li and $^{26\!}$Al within a network
of charged-particle reactions.

The focus of this article is to review the $\nu$ process in the light
of an improved understanding of neutrino properties in core-collapse
supernovae and with an improved set of neutrino-nucleus cross sections covering all
nuclei in the reaction network
to explore in particular the impact of the $\nu$ process on the
production of long-lived radioactive nuclei of interest to gamma-ray
astronomy. Previous investigations of nucleosynthesis by
neutrino-induced reactions have been based on stellar simulations
using various hydrodynamical
models~\citep{Woosley.Hartmann.ea:1990,Heger.Kolbe.ea:2005,Limongi.Chieffi:2006}
and neutrino-nucleus cross section data which were restricted to a set
of key nuclei, especially those which are quite abundant in outer burning
shells, and to a limited number of decay channels. Furthermore, the
simulations adopted supernova neutrino energy spectra, described by
Fermi-Dirac distributions with chemical potential $\mu=0$ and
temperature $T_{\!\nu}$, which were appropriate at the time the studies
were performed.  They used $T_{\nu_{\mathrm{e}},\bar{\nu}_{\mathrm{e}}}=4$-5~MeV for electron
(anti)neutrinos, corresponding to average energies,
$\langle E_\nu\rangle = 3.15\, T_{\!\nu}$, between 12~MeV and
16~MeV~\citep{Woosley.Hartmann.ea:1990,Heger.Kolbe.ea:2005} and
$T_{\nu_{\mu,\tau}}=5$-10~MeV~\citep{Woosley.Hartmann.ea:1990,
  Timmes.Woosley.ea:1995,Heger.Kolbe.ea:2005} for muon and tau
neutrinos as well as for the corresponding anti-neutrinos,
corresponding to average energies between 16~MeV and 32~MeV.

We improve these simulations in two relevant aspects.  Firstly, we
have derived a complete set of partial differential cross sections for
neutrino-induced charged- and neutral-current reactions for nuclei
with charge numbers $Z < 76$ considering various single- and
multi-particle decay channels. For several key reactions, we have
derived the cross sections either directly from experimental data or
from  shell model calculations which is the most accurate theoretical tool to
describe low-energy neutrino-nucleus reactions \citep{Langanke.Martinez:2003,Balasi.Langanke.ea:2015}. The relevant cross
sections are provided in the supplemental material. Secondly, the more
realistic treatment of neutrino transport in recent supernova
simulations~\citep{Fischer.Whitehouse.ea:2010,Huedepohl.Mueller.ea:2010,Martinez-Pinedo.Fischer.ea:2012,Martinez-Pinedo.Fischer.Huther:2014,Mirizzi.Tamborra.ea:2016}
yield spectra for all neutrino families which are noticeably shifted
to lower energies.  This reduces the neutrino-nucleus cross sections and
in particular particle spallation cross sections for neutral-current
reactions which are very sensitive to the tail of the neutrino
spectra due to the relatively high particle separation thresholds involved. Our choice of neutrino temperatures denoted ``\emph{low energies}''
is $T_{\!\nu_{\mathrm{e}}}= 2.8$~MeV ($\langle E_{\nu_{\mathrm{e}}}\rangle = 9$~MeV),
$T_{\!\bar{\nu}_{\mathrm{e}},\nu_{\mu,\tau}}= 4$~MeV 
($\langle E_{\bar{\nu}_{\mathrm{e}},\nu_{\mu,\tau}}\rangle = 12$~MeV), in agreement
with recent
simulations~\citep{Huedepohl.Mueller.ea:2010,Martinez-Pinedo.Fischer.ea:2012,Martinez-Pinedo.Fischer.Huther:2014,Mirizzi.Tamborra.ea:2016}. To
compare with previous neutrino nucleosynthesis
studies~\citep{Woosley.Hartmann.ea:1990,Heger.Kolbe.ea:2005} we have
also performed our calculations using the following set of neutrino
temperatures: $T_{\!\nu_{\mathrm{e}}}=4$~MeV
($\langle E_{\nu_{\mathrm{e}}}\rangle = 12$~MeV), $T_{\!\bar{\nu}_{\mathrm{e}}}=5$~MeV
($\langle E_{\bar{\nu}_{\mathrm{e}}}\rangle = 15.8$~MeV), and
$T_{\!\nu_{\mu,\tau}}=6.0$~MeV
($\langle E_{\nu_{\mu,\tau}}\rangle = 19$~MeV); that we denote as
``\emph{high energies}'' along the manuscript.

The paper is organized as follows: In \S\ref{sec:rates} we discuss our
calculations of neutrino-nucleus reactions with a particular focus on
the cases for which experimental data are available.  After a brief
description of the supernova model in \S\ref{sec:model}, we
discuss our results for the stable nuclei in \S\ref{sec:stable}
and then the impact on radioactive nuclei in \S\ref{sec:radio}. 
Finally, we conclude in \S\ref{sec:conclusions}.

\section{Neutrino-nucleus cross sections}
\label{sec:rates}

We have calculated partial differential neutrino-nucleus cross
sections globally for nuclei with $Z < 76$ based on a two-step
strategy~\citep{Kolbe.Langanke.ea:1992}: \textsl{i)} the neutrino-induced
nuclear excitation cross sections to a final state at energy $E$ have
been calculated within the Random Phase Approximation
following~\citet{Kolbe.Langanke.ea:2003} and considering multipole
transitions up to order four. The single particle energies were
adopted from an appropriate Woods-Saxon parametrization, adjusted to
reproduce the proton and neutron thresholds and to account for the
energies of the Isobaric Analog State and the leading giant
resonances. \textsl{ii)} The decay probabilities of the excited nuclear levels
have been derived within the statistical model. At low excitation
energies we use the Modified Smoker code~\citep{Loens:2010} which
considers experimentally known states and their properties explicitly
and then matches the experimental spectrum to a level density.  The
code is restricted to treat single-particle decays.  To allow for
multi-particle decay, which becomes relevant at modest excitation
energies or in nuclei with large neutron excess and hence small
separation energies, we have adopted the ABLA
code~\citep{Kelic.Valentina.Schmidt:2009} at higher excitation
energies, which has been validated to properly describe
multi-particle decays and fission. The results of the two statistical
model codes have been smoothly matched at moderate energies above the
single-particle thresholds.

Using cross sections for neutrino nucleus reactions based on Random
Phase Approximation (RPA) is justified by the relatively high energy of
incoming neutrinos that are large compared to the energy scale of nuclear
excitations. In this case collective excitations dominate and in RPA
the centroid of the strength of collective excitations are reproduced
quite reliably without the need of a precise reproduction of the detailed fragmentation.

\begin{figure}[htb]
  \includegraphics[width=0.48\textwidth]{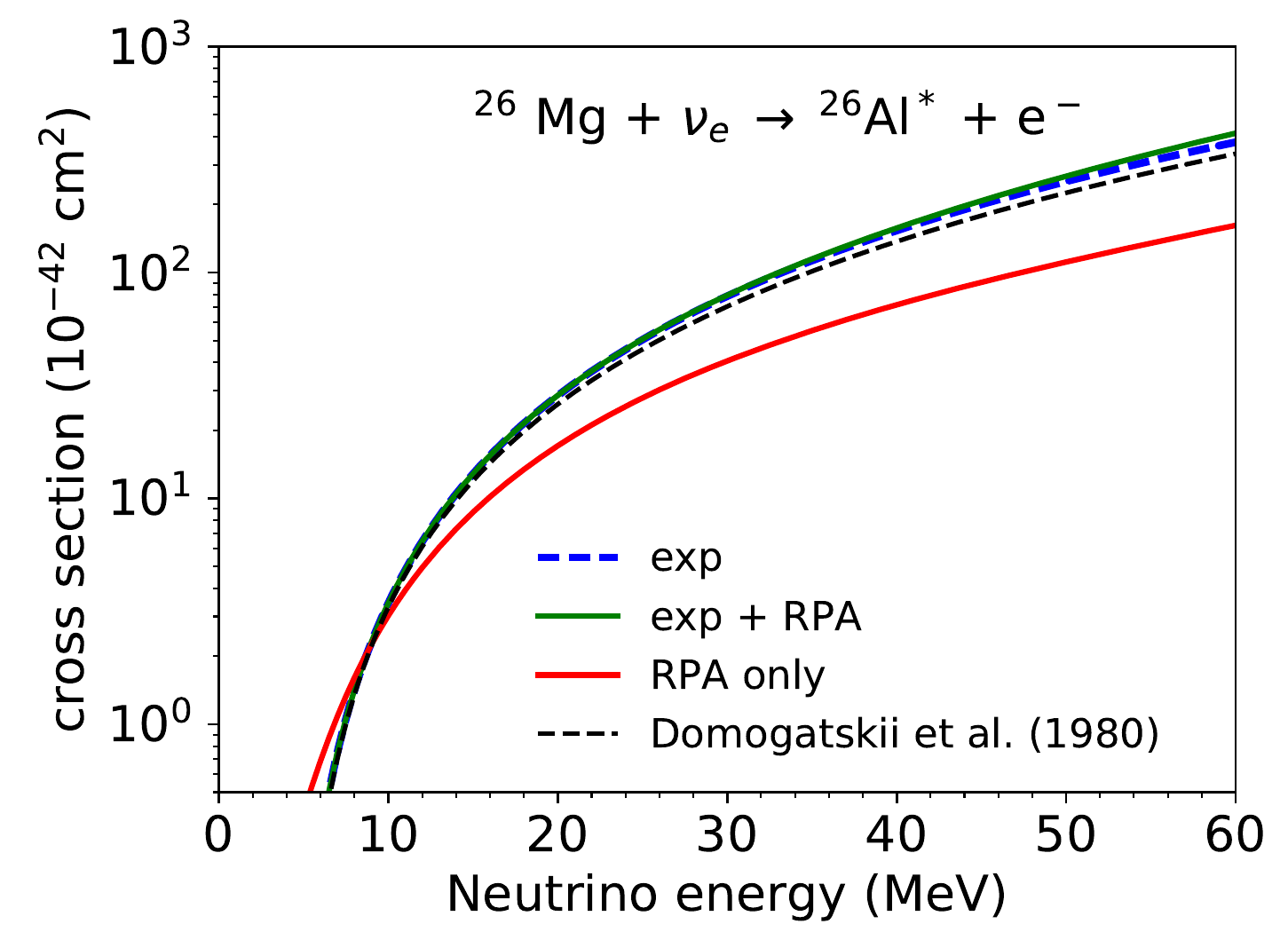}
  \caption{Cross section for the reaction
    $^{26}$Mg($\nu_{\mathrm{e}}$,$\mathrm{e}^-$)${}^{26}\!$Al based on the experimentally
    determined strength only (exp) and also with additional strength
    from calculated transitions at higher energies (exp + RPA). For
    comparison the theoretical results based on RPA and the estimate
    of~\citet{Domogatskii.Nadezhin:1980} are also
    shown.\label{fig:nuemg26}}  
\end{figure}

A particular relevant nucleus for gamma-ray astronomy is $^{26\!}$Al
that is produced in the $\nu$ process by the reaction
$^{26}$Mg$(\nu_{\mathrm{e}},\mathrm{e}^-){}^{26}\!$Al and will be
discussed in detail in \S\ref{sec:radio}.  This reaction has been
considered previously by~\citet{Domogatskii.Nadezhin:1980} where the
cross section was estimated considering the Fermi contribution to the
isobaric analog state and Gamow-Teller transitions from the $\beta^+$
decay of $^{26}$Si, the mirror nucleus to
$^{26}$Mg. \citet{Zegers.Akimune.ea:2006} have extracted the B(GT)
strength distribution based on $^{26}$Mg$(^3\mathrm{He},\mathrm{t})$
charge exchange reactions with high resolution. This has the advantage
of not having the energy limitations of beta-decay and allows to
determine transitions to all relevant states in $^{26}\!$Al. Based on
these data and the known Fermi transition to the isobaric analog state
we have calculated the cross sections for
$^{26}$Mg$(\nu_\mathrm{e},\mathrm{e}^{-})^{26}\!$Al. In order to
account for additional contributions from forbidden transitions
relevant at high neutrino energies, the cross section has been
supplemented by RPA calculations. As can be seen in
Figure~\ref{fig:nuemg26}, the contribution to the cross section from
forbidden transitions is rather small.  Also shown is the estimate
of~\citet{Domogatskii.Nadezhin:1980} and the purely theoretical RPA
cross section.  With appropriate quenching factors RPA calculation
have been shown to give good estimates for the for the centroid of the
B(GT) strength and to fulfill sum rules. It does not reproduce the
fragmentation of the distribution very well
\citep{Kolbe.Langanke.ea:2003,Balasi.Langanke.ea:2015} which is
important for reactions on intermediate mass nuclei at low energies
that are dominated by a few individual transitions.  The
charged-current reaction
$^{26}$Mg$(\nu_\mathrm{e},\mathrm{e}^-)^{26}\!$Al is particularly
difficult to treat in the RPA because it is sensitive to the
distribution of the strength at low energies. Furthermore, for nuclei
that are close to $N=Z$ the Gamow-Teller strength is not determined by
the Ikeda sum rule as it is the case of nuclei with moderate and large
neutron excess.  Figure~\ref{fig:nuemg26} also shows the cross
sections based on the RPA calculations and we find that it is 60\%
smaller than the one based on data for the relevant neutrino energies
This discrepancy exceeds the typical uncertainty that are expected
from RPA neutrino cross section
calculations~\citep{Balasi.Langanke.ea:2015}. For neutral-current
interactions only reactions that lead to particle emission are
relevant for nucleosynthesis and electron antineutrinos are expected
to have higher energies. For such reactions, collective excitations at
higher energies become more relevant and the cross sections are less
sensitive to the details of the strength distribution.  Therefore, we
expect the values for such reactions to be even more accurate.
\citet{Caurier.Langanke.ea:1999} and others have shown that the shell
model can describe the details of the B(GT) strength distribution with
high accuracy \citep[see
also][]{Balasi.Langanke.ea:2015}.  We use results from
such calculations wherever available.

The radioactive nucleus $^{36}$Cl is of some particular interest
because it has been found in material from the early solar
system~\citep{Murty.Goswami.Shukolyukov:1997}. The reaction
$^{36}$S$(\nu_\mathrm{e},\mathrm{e}^-)^{36}$Cl can contribute to the
production of $^{36}$Cl in supernovae. We have determined the cross
section by combining the data from the mirror $\beta^+$ decay of
$^{36}$Ca and Gamow-Teller strength determined by shell-model
calculations using the USDB interaction~\citep{Brown.Richter:2006}
with forbidden contributions determined from the RPA calculations. A
secondary production channel for the production of $^{36}$Cl is
provided by the reaction
$^{36}\!$Ar$(\bar{\nu}_\mathrm{e},\mathrm{e}^-)^{36}$Cl that we have
determined by combining shell-model calculations for the Gamow-Teller
strength and RPA for the forbidden transitions.\\ The reaction
$^{22}$Ne($\nu_\mathrm{e},\mathrm{e}^-)^{22}$Na is relevant for the
production of the radioactive isotope $^{22}$Na and has been
determined by using data from the $\beta^+$ decay of the mirror
nucleus $^{22}$Mg combined with shell-model calculations for the
Gamow-Teller strength and RPA calculations for the forbidden strength.

Other relevant cross sections are taken from the
literature. Cross-sections for $^{4}$He are taken
from~\citet{Gazit.Barnea:2007}. 
For $^{12}$C($\nu$,$\nu$' $^3$He)$^{9}$Be,$^{12}$C($\nu$,$\nu^\prime$ $^4$He $^3$H
p)$^{4}$He, and $^{12}$C($\nu$,$\nu^\prime$  $^4$He $^3$He n)$^{4}$He 
the values used
by~\cite{Woosley.Hartmann.ea:1990} are adopted.  One and two proton and neutron
emission from $^{12}$C,$^{14}$N,$^{16,18}$O, $^{20,22}$Ne, $^{24,26}$Mg, and
$^{26,28}$Si follow the approach discussed by~\citet{Heger.Kolbe.ea:2005}.  The
cross sections on $^{20}$Ne is based on charge-exchange
data~\citep{Anderson.Tamimi.ea:1991} now extended to both neutral-current and
charged-current cross sections. 
Cross sections for $\nu_\mathrm{e}$ absorption
on $^{138}$Ba and $^{180}$Hf are based on measured Gamow-Teller
strengths~\citep{Byelikov.Adachi.ea:2007} with branching ratios for
particle emission based on a statistical model \citep{Loens:2010}.

In addition to the updated $\nu$-induced reaction rates our work
includes recent updates of thermonuclear reactions rates, in
particular proton-, neutron- and alpha capture rates as contained in
the most recent release of the JINA REACLIB reaction rate library,
Version~2.2 \citep{Cyburt.Amthor.ea:2010} which contains important
improvements for example on proton induced
reactions~\citep{Iliadis.DAuria.ea:2001} and neutron capture
rates~\citep{Dillmann.Szuecs.ea:2014}. Neutron capture rates are
particularly important to determine the final abundances of $^{138}$La
and $^{180}$Ta.  While we use the updated rates from KADoNiS v0.3 for
$^{179,180}$Ta$(n,\gamma)$, we revert to the values by \citet{Rauscher.Thielemann.ea:2000}
for $^{137,138}$La$(\mathrm{n},\gamma)$ for reasons of consistency as explained in \S\ref{sec:lata}.

\section{Supernova model and nuclear reaction network}
\label{sec:model}
The $\nu$ process operates before, during and after the shock wave
reaches the different regions of the star. The evolution of the
shock-wave passing through the outer layers of the star is calculated
using the implicit hydrodynamics package KEPLER
\citep{Weaver.Zimmerman.Woosley:1978,Woosley.Weaver:1995,
Woosley.Heger.Weaver:2002}.
In this framework, the explosion is driven by a piston that is
positioned at the edge of the Fe core and the trajectory of the piston
is adjusted to achieve an explosion energy of $1.2\times 10^{51}$~erg
for all the models we have computed.  We use supernova progenitors
from a set that has been evolved in the same
numerical framework as discussed by \cite{Rauscher.Heger.ea:2002}, spanning initial
masses between 13-30~M$_\odot$. It is unclear which of the explored
models would explode self-consistently and how the explosion energy
and amount of fallback depend on progenitor mass and
structure~\citep{Woosley.Weaver:1995,Horiuchi.Nakamura.ea:2014,
Sukhbold.Woosley:2014,Ertl.Janka.ea:2016}.  Taking the same
explosion energy for all the models probably also affects the
systematics with respect to the progenitor mass.  The
progenitor-explosion connection, however, is still an open question and active field of research \citep{Mueller.ea:2016,Sukhbold.ea:2018}.

The progenitor models we study here have been
affected by the coding error affecting the neutrino loss rates
reported by \citet{Sukhbold.ea:2018} that affect the progenitor
structure, in particular the innermost regions. However, the $\nu$
process operates mostly regions beyond the O/Ne shell which we are not
significantly affected by this error.

Since our results are based on one-dimensional calculations the
results depend on the choice of the mass cut, which determines the
amount of the material that is accreted onto the central object
(fallback) and determines how much of the innermost part of the star can
be successfully ejected by the explosion. Since the $\nu$ process
mainly operates in outer regions of the stellar mantle the results
should not be affected significantly by fallback. Nevertheless,
fallback may trigger the formation of a black hole resulting in a
sudden end of neutrino emission~\citep{Fischer.Whitehouse.ea:2009}. This possibility is neglected in
our calculations.

For the sake of comparison with previous studies and since neutrino energies
and luminosities from self-consistent explosion simulations are still rare and
the quantitative relations to the progenitor model are not established yet  \citep{Mueller.ea:2016,Sukhbold.ea:2018}
 we model
the neutrino emission with a exponentially decreasing luminosity
$L_{\nu}=L_{0}\, \text{exp}(-t/\tau_\nu)$ with $\tau_\nu=3$~s and
$L_{0}$ chosen to result in a total energy of $3 \times 10^{53}$~erg
emitted as neutrinos and distributed equally over the six neutrino
flavors. 
\begin{table*}[htb]
\caption{Production factors relative to solar abundances from~\citet{Lodders:2003}, normalized to $^{16}$O
production. Shown are the results obtained without neutrino, with
our choice of neutrino temperatures (``Low energies''), and with
the choice of~\citet{Heger.Kolbe.ea:2005} (``High energies''). For each
set of energies, the results are also shown when only charged current reactions (induced by electron flavor neutrinos)
are considered and when only neutral current reactions are considered.
\label{tab:prodfac}}
\begin{ruledtabular}
\begin{tabular}{llcccccc}
Nucleus& no $\nu$ & \multicolumn{3}{c}{Low
 energies\footnote{$T_{\nu_\mathrm{e}}=  2.8$~MeV,
                    $T_{\bar{\nu}_\mathrm{e}}=T_{\nu_{\mu,\tau}}= 4.0$~MeV}} &
    \multicolumn{3}{c}{High
                                                                     energies\footnote{$T_{\nu_\mathrm{e}}=4.0$~MeV,$T_{\bar{\nu}_\mathrm{e}}=5.0$~MeV,     
    $T_{\nu_{\mu,\tau}}=6.0$~MeV}}\\  \cline{3-5}\cline{6-8}
	 &      & with $\nu$ & only charged current & only neutral current  &  
		  with $\nu$ & only charged current & only neutral current \\ \hline

$^{7}$Li   &  0.002  &  0.04  &  0.01  &  0.03  &  0.58  &  0.05  &  0.57  \\
$^{11}$B   &  0.01  &  0.31  &  0.17  &  0.21  &  1.57  &  0.58  &  1.31  \\
$^{15}$N   &  0.06  &  0.09  &  0.08  &  0.08  &  0.16  &  0.10  &  0.15  \\
$^{19}$F   &  0.13  &  0.18  &  0.14  &  0.16  &  0.29  &  0.17  &  0.26  \\
$^{138}$La &  0.16  &  0.46  &  0.44  &  0.18  &  0.77  &  0.73  &  0.22  \\
$^{180}$Ta\footnote{Assuming that 35\% survives in the
long-lived isomeric state~\citep{Mohr.Kaeppeler.Gallino:2007}}  
	 & 0.20  &  0.49  &  0.48  &  0.24  &  0.84  &  0.80  &  0.33  \\
\end{tabular}
\end{ruledtabular}
\end{table*}
Abundances are evolved using a nuclear reaction network including 1988 species
up to $2.5\times 10^4$~s after bounce when the most short-lived nuclei have
already decayed and those potentially interesting for observations remain.  If
not stated otherwise, mass fractions of radioactive nuclei quoted here have
been extracted at this time. Nuclear reactions are switched off, once the
temperature drops below $10^{7}$ K.  However beta-decays and neutrino reactions
are followed till the end of the calculation.  The size of the nuclear reaction
network matches the co-processing network that was employed in the calculations
of the stellar evolution of the progenitor models
\citep{Woosley.Heger.Weaver:2002}.  Therefore, any effects from the $s$~process
during stellar evolution are included. This is particularly important for the
nucleosynthesis of the heaviest species: $^{92}$Nb, $^{98}$Tc, $^{138}$La and
$^{180}$Ta.

\section{Stable isotopes}
\label{sec:stable}
The typical nuclei that are sensitive to neutrino nucleosynthesis are
$^{7}$Li, $^{11}$B, $^{15}$N, $^{19}$F, $^{138}$La, and
$^{180}$Ta~\citep{Woosley.Hartmann.ea:1990,Heger.Kolbe.ea:2005}, all
of which are observed in the solar system, but are not produced in
sufficient amount by nucleosynthesis calculations without including
neutrino interactions. Neutrino nucleosynthesis pushes the averaged
production factors of those nuclei closer to the solar system
values. Table~\ref{tab:prodfac} shows the production factors relative
to $^{16}$O averaged over our set of progenitors weighted with a
Salpeter initial mass function (IMF) with
$dN_*/dm_* \propto m_*^{-1.35}$. The production factor for a species
$A$ is defined as
$P_A=(X^*_A/X_A^\odot)/(
  X^*_{^{16}\text{O}}/X_{^{16}\text{O}}^\odot)$. We find that due to the
reduction of the $\nu$ energies the effect of the $\nu$ process is
diminished which solves the problem of the slight overproduction of
$^{11}$B.

\begin{figure}[htb]
  \includegraphics[width=\linewidth]{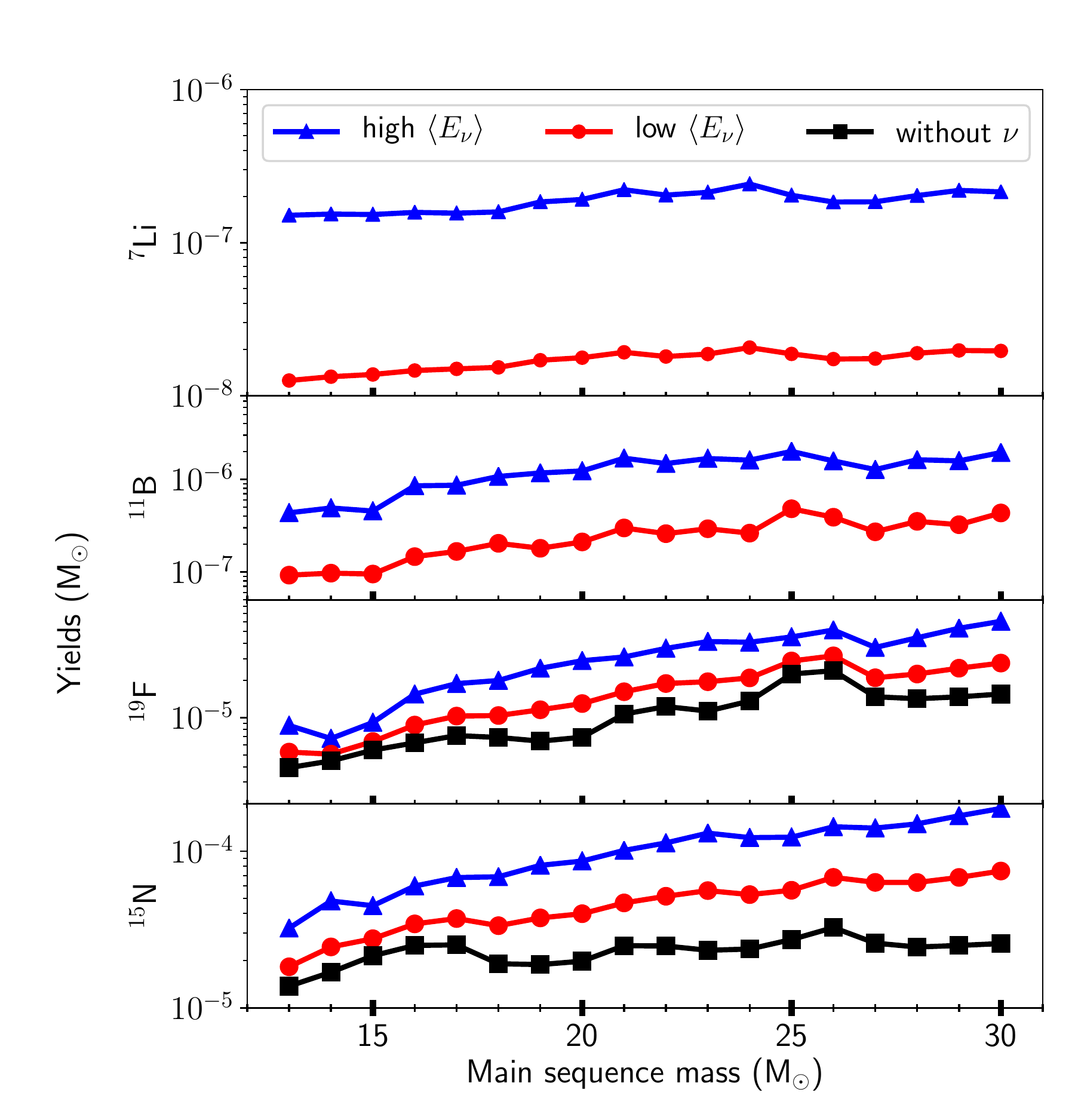}
  \caption{Total yields for the nuclei with largest contributions from
    the $\nu$ process for the range of progenitors studied
    here. Averaged production factors are summarized in
    Table~\ref{tab:prodfac}. Note that the scale of the y axis is
    different for each panel. The production of $^7$Li and $^{11}$B without neutrinos is
      negligible and not shown in the figure.\label{fig:stable_yields}}
\end{figure}

Figure \ref{fig:stable_yields} shows the dependence of the total
yields of the five $\nu$-process isotopes shown in
Table~\ref{tab:prodfac} on the initial mass of the stellar
model. Stellar structure affects the $\nu$ process by three major
aspects. Firstly, as a secondary process the $\nu$ process
predominantly 
operates on abundant seed nuclei and the composition therefore
determines where the process can occur. Secondly, the stellar density
and temperature profiles determines how strong the supernova shock
affects the regions where the $\nu$ process seeds are
located. Finally, the stellar model determines the distance of the
regions of interest from the proto-neutron star and hence the
intensity of the neutrino fluxes.

Due to the complex interplay of nuclear
burning, convection and hydrostatic adjustment that governs stellar evolution
monotonous trends with respect to the initial mass are not expected.
Still, Figure~\ref{fig:stable_yields} shows that the relative
enhancement for the $\nu$-process nuclei and in particular for the
light elements Li and B are quite robust with respect to the
progenitor. In general, the $\nu$-process contributions tend to have a smoothing effect
on variations with initial mass which we also find for the radioactive
nuclei discussed in \S\ref{sec:radio}.
In the following we will discuss the production of $^7$Li, $^{11}$B, 
$^{19}$F, $^{15}$N, $^{138}$La and $^{180}$Ta in more detail.

\subsection{The light nuclei $^7$\textup{Li} and $^{11}$\textup{B}}
\label{sec:light-elements-li}
The light elements $^7$Li and $^{11}$B are present in the solar system with abundances of $1.5\times 10^{-9}$ and  $4\times 10^{-10}$ \citep{Lodders:2003}.
Since these nuclei are easily destroyed by charged particle reactions there is no stellar production mechanism and the origin of theses abundances is a long-standing
problem. As discussed \citep[e.g., by][]{Prantzos:2007} irradiation by galactic cosmic rays (GCR) is a promising 
scenario but also associated with large uncertainties. Most likely a combination of the contributions from GCR irradiation and the $\nu$~process in 
core-collapse supernovae is required to explain the solar abundance as pointed out by \citet{Prantzos:2012} and \citet{Austin.West.Heger:2014}.

\begin{figure}[htb]
  \includegraphics[width=\linewidth]{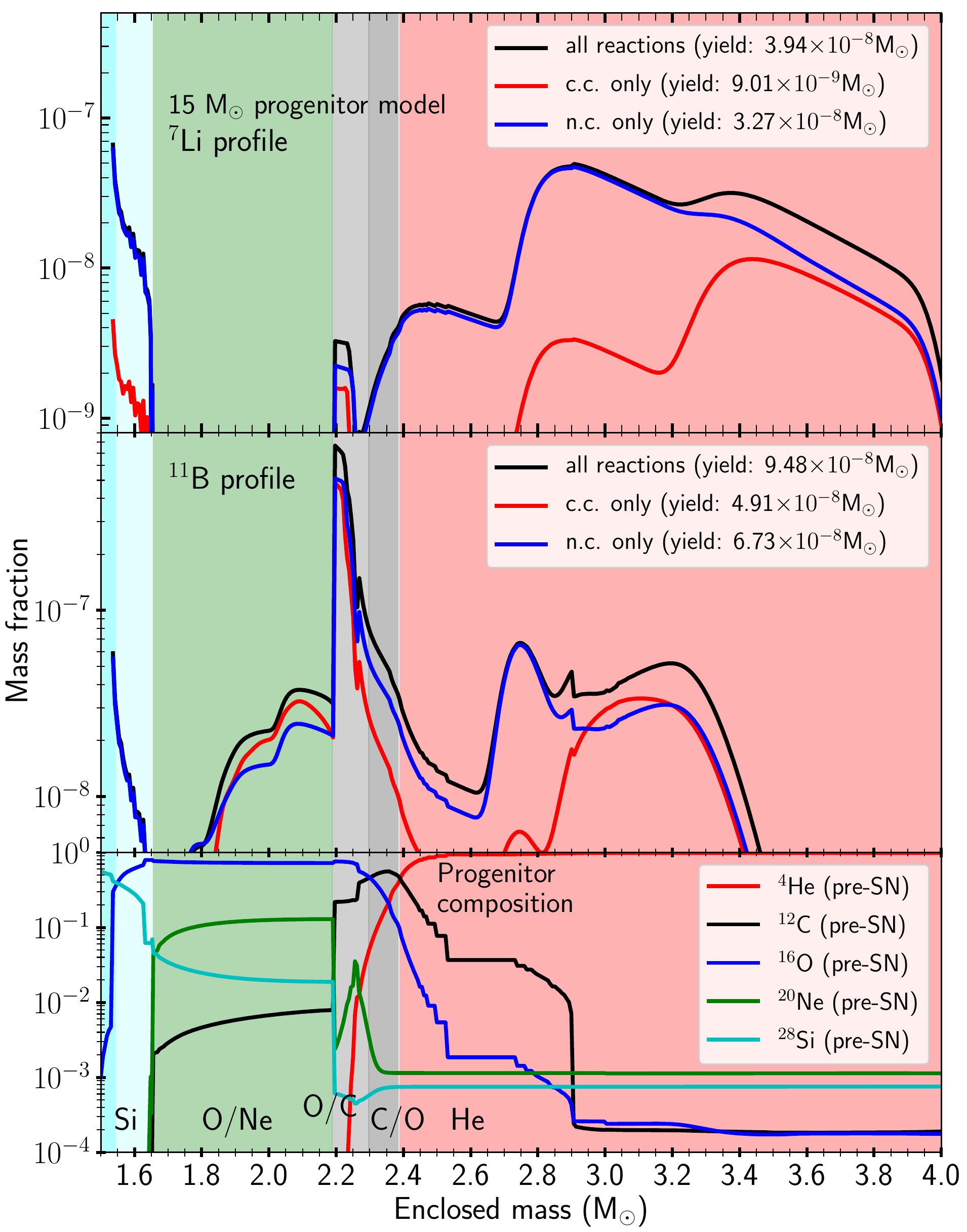}
  \caption{Profiles of $^{7}$Li and $^{11}$B mass fractions for the
    15~M$_\odot$ using the updated low neutrino energies model with
    solar metallicity. $^{7}$Li is mostly made at the base of the He
    shell by the reaction sequence described in the text and $^{11}$B
    is produced in the thin C layer. Also shown are results when only
    charged current (c.c.) or only neutral current (n.c.)  reactions
    are taken into account. This shows that electron flavor
    neutrinos contribute to as similar extent as the heavy flavor
    neutrinos for . The colored regions denote the different
    stellar regions as labeled in the bottom panel: Si shell (cyan), O
    shell (light cyan), O/Ne shell (green), O/C shell (grey), and He
    shell (red) \label{fig:LiB_s15}}
\end{figure}

To illustrate how the $\nu$~process can produce the light elements the upper
panel of Figure~\ref{fig:LiB_s15} shows the $^7$Li mass fraction as a function
of the mass coordinate for a 15~M$_\odot$ progenitor model for the set of low
neutrino energies as defined above.  In order to disentangle the impact of
electron type (anti)neutrinos and heavy flavor neutrinos results from
calculations in which the neutrino reactions either only for charged- (c.c.) or
only for neutral current (n.c.) process have been included are also shown.

At the base of the He-shell the neutral current neutrino-interactions
$^{4}$He$(\nu,\nu'p)$ and $^{4}$He$(\nu,\nu'n)$ contribute to produce
$^{7}$Li by the reactions
$^{3}$He($\alpha$,$\gamma$)$^{7}$Be($\beta^+$)$^{7}$Li and $^{11}$B
via $^{3}$H($\alpha$,$\gamma$)$^{7}$Li($\alpha$,$\gamma$)$^{11}$B. The
same reaction chains operate on the $\alpha$ rich freeze-out operating
at the base of the Si shell in the most inner supernova ejecta near to
the mass cut where the material is subject to the most intense
neutrino irradiation. This can be seen in Figure~\ref{fig:LiB_s15} for
the 15 M$_\odot$ progenitor model, where a noticeable mass fraction of
$^{7}$Li and $^{11}$B are produced right at the edge of the mass
cut. The contribution from this region to the total yield is of the order of a few
percent and thus negligible in the models considered here, but it
depends on the choice of the mass cut in parametric 1D simulations and
is sensitive to details of the explosion dynamics. If the material is
kept close to the neutron star for a longer time due to turbulent
convection and is later ejected, the contribution of this region to
the production of $^{7}$Li and $^{11}$B could be larger.  A final
answer to the role of the $\nu$ process in this region therefore
requires to take into account multi-D effects from self-consistent
supernova simulations.

The bulk of $^{11}$B is produced in the thin C shell, see middle panel
of Figure~\ref{fig:LiB_s15}, by neutral-current spallation reactions
on $^{12}$C, i.e., $^{12}$C$(\nu,\nu'n/p)$, that mostly produce
$^{11}$C that decays later with a half-life of about 20 minutes to
$^{11}$B.  The charged current reactions
$^{12}$C$(\nu_e,e^{-}p)^{11}$C and
$^{12}$C$(\bar{\nu}_e,e^{+}n)^{11}$B contribute almost as much as the
neutral current for the low energies.  However, for the high
energies the neutral current clearly dominates.  This is also visible
in the averaged values in Table \ref{tab:prodfac}, where the
production factors for the calculations with only neutral- and charged
current reactions are also shown. It illustrates the increased
importance of charged current processes for the ``low energies'' case.
Figure~\ref{fig:LiB_s15} also shows that there
is a minor contribution to
$^{11}$B in the O/Ne shell which is due to $^{16}$O$(\nu,\nu' \alpha p)$.  Such
multi-particle emission channels have not been included in previous studies but
are now taken into account for all nuclei in the reaction network.

The production of $^{7}$Li and $^{11}$B requires the knockout of
protons and neutrons from tightly bound $^4$He, $^{12}$C and $^{16}$O
by high energy neutrinos from the tail of the
distribution. Consequently, the shift of the neutrino spectra to lower
energies has a significant impact on the production of these light
elements.
Due to the sensitivity of $^{7}$Li and $^{11}$B to the neutrino energies
\cite{Yoshida.Kajino.ea:2005} have suggested that the energies of
$\nu_{\mu,\tau}$ and $\bar{\nu}_{\mu,\tau}$ can be constrained requiring a good
reproduction of the solar abundance of $^{11}$B considering contributions from
both cosmic-rays and $\nu$ process.  We use updated neutrino cross sections for
reactions on $^4$He from~\cite{Gazit.Barnea:2007} that are slightly larger than
those previously used, giving an increase in the production of $^{7}$Li and
$^{11}$B compared to the yields presented by~\cite{Heger.Kolbe.ea:2005} when we
use the same energies (``high energies'' case).  The same cross sections for
$^4$He and also comparably low energies have also been used in a recent study
by \citet{Banerjee.Qian.ea:2016} about the production of radioactive $^{10}$Be
for a supernova model of a 11.8~M$_\odot$ progenitor.  $^{10}$Be is mostly
produced by the two proton emission channel $^{12}$C$(\nu,\nu' p p)^{10}$Be in
the O/C shell and for the models studied here we obtain an average yield of
$3.9\times 10^{-11}$~M$_\odot$ for the low neutrino energies and  $2.5\times
10^{-10}$~M$_\odot$ for the high energies. Even with
the high neutrino energies the $\nu$~process does not produce enough  $^{10}$B
and  $^{6}$Li  to explain the solar system values. While the solar system
value for $^{11}$B/$^{10}$B is around $4$, the average value we get from our
calculations is $^{11}\mathrm{B}/^{10}\mathrm{B} \approx 100$. Similarly we find
$^{7}\mathrm{Li}/^{6}\mathrm{Li}\approx 200$ compared to a solar system value of $12$.
Therefore, the abundances of $^{10}$B and $^{6}$Li require a contribution from
cosmic ray spallation~\citep[e.g.,][]{Prantzos:2007} which also contributes to
$^{11}$B and $^{7}$Li.  Based on models for the production $^{11}$B and
$^{10}$B by galactic cosmic rays (GCR), \cite{Austin.West.Heger:2014} have
estimated that the $\nu$ process must produce about $42\pm4$\% of the solar
$^{11}$B which is consistent with our estimate based on the ``low energies''
case. 

We find that with realistic neutrino energies the production of $^7$Li
by the $\nu$ process is negligible. This is consistent with the observation of
the Lithium ``Spite
plateau''~\citep{Spite.Spite:1982,Sbordone.Bonifacio.ea:2010} in metal-poor
stars in the metallicity range $-3.0 \lesssim \text{[Fe/H]} \lesssim -1.5$.
The top panels of Figure \ref{fig:stable_yields} show that the neutrino
enhanced yields of $^{7}$Li and $^{11}$B are not as sensitive to the progenitor
model as, e.g., $^{19}$F. This is
because in the lower mass stars the relevant zones tend to be closer to the
proto- neutron star which compensates for the overall narrower burning shells
that contain smaller amounts of relevant seed nuclei. Since the production of
$^{16}$O increases with progenitor mass that also means that the production
factor normalized to $^{16}$O significantly increases towards the low mass end
of the progenitor range we studied. Hence uncertainties in the initial mass
function will also play an important role since the weight given to the low
mass stars is crucial for the average production factor.

\subsection{$^{15}$\textup{N}, and $^{19}$\textup{F}}

The $\nu$ process can contribute to the production of $^{15}$N and to $^{19}$F
in the O/Ne and O/C shells mostly via the neutral current spallation of protons
or neutrons $^{16}$O$(\nu,\nu' p/n)$ and $^{20}$Ne$(\nu,\nu' p/n)$
respectively, since $^{15}$O as well as $^{19}$Ne quickly decay to $^{15}$N and
to $^{19}$F respectively.  Even the charged current
reactions $^{16}$O/$^{20}$Ne$(\nu_e,e^{-} p)$ and
$^{16}$O/$^{20}$Ne$(\bar{\nu}_e,e^{+} n)$ finally contribute to
$^{15}$N and to $^{19}$F. 
When we take into account the harder spectrum for the heavy flavor neutrinos
we find that
the spectrally averaged cross section for the sum of the two charged current channels is a factor $10$ smaller than the
combined neutral current channels for the higher neutrino
energies. For the lower energies the charged current contribution
is now smaller by only a factor $3$.

\citet{Heger.Kolbe.ea:2005} have already argued that this mechanism
can probably not account for the entire solar abundance of $^{19}$F
and can only produce small amounts of $^{15}$N. 
Table \ref{tab:prodfac} shows that with the low neutrino energies the
averaged $^{19}$F yield is increased by 30\% but still only reaches
a production factor of $0.2$ and with high energies it is less than $0.3$.
This is in agreement with the conjecture that
the $\nu$~process in core-collapse supernovae is not the only source of 
$^{19}$F which is
supported by recent observational evidence.
Spectral analysis of
nearby stars do not show a distinct correlation between O and F
abundances that would be expected if supernovae were the main source
for Florine~\citep{Joensson.Ryde.ea:2017}.  Galactic chemical evolution
models~\citep{Renda.Fenner.ea:2004,Kobayashi.Izutani.ea:2011},
still attribute a significant component of the galactic $^{19}$F
inventory to core collapse supernovae in combination with
asymptotic giant branch (AGB) and Wolf-Rayet stars. 

The lower panel of Figure \ref{fig:stable_yields} shows that the $^{19}$F yield 
exhibits a relatively large sensitivity to the progenitor model. Indeed, the 
mechanism behind the production depends significantly on the mass of
the star. 

For the 15~M$_\odot$ model the supernova shock alone, i.e., without
neutrinos, increases the pre-supernova $^{19}$F content of
$4.3\times 10^{-6}$~M$_\odot$ to a yield of
$5.5\times 10^{-6}$~M$_\odot$, corresponding to a production factor of
0.15.  Neutrinos increase the production factor to 0.20 or 0.28 for
low and high energies respectively.  The thermonuclear component is
mainly due to the reaction sequence
$^{18}$O$(p,\alpha)^{15}$N$(\alpha,\gamma)^{19}$F operating on
$^{18}$O at the lower edge of the He-shell where post shock
temperatures reaches 0.67~GK at densities of up to 1500~g~cm$^{-3}$.
This requires an episode of convection to mix the $^{18}$O from the
CNO cycle down to the bottom of the He-shell where the peak
temperature in the shock will be optimal.  For the least massive star,
the 13 M$_\odot$ model $^{18}$O remains concentrated in a narrow
region where the peak temperature reaches less than 0.5 GK and as a
results the shock heating does not really play a role for the $^{19}$F
yield without neutrinos which here results almost entirely from the
pre explosive hydrostatic burning and gives a production factor of
0.23.  Including the $\nu$~process in this model however gives the
highest production factor among the models studied here of 0.27 for
the low energies and 0.37 for high energies.

\begin{figure}[htb]
\includegraphics[width=\linewidth]{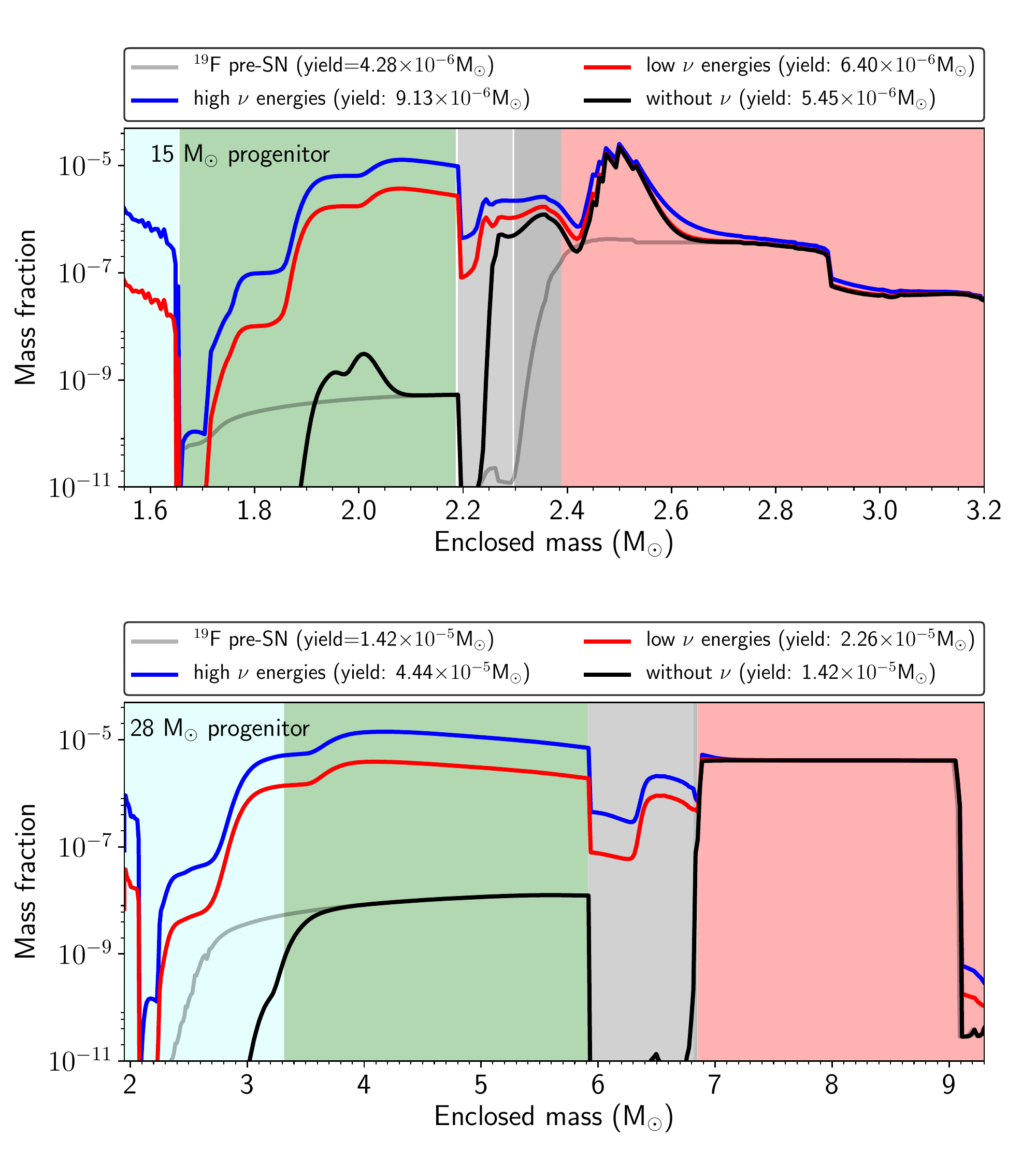}
\caption{Mass fraction of $^{19}$F for the  15~M$_\odot$ model (upper panel)
and the 28~M$_\odot$ progenitor (lower panel) representative for the lower
and upper ends of the range of masses we explore. Shown are the pre-SN mass
fractions as well as the final mass fractions without neutrinos, with the
updated set of low neutrino energies and the high energies.  While
the supernova shock leads to a peak in the mass fraction around mass coordinate
2.5~M$_\odot$, the $\nu$~process is much more prominent for the more massive
model.  The background colors indicate the compositionally differing shells of
the progenitor star as shown in detail at the bottom of Figure
\ref{fig:LiB_s15} \label{fig:f19_examples}}
\end{figure}

The profile of the $^{19}$F mass fraction for the 15 M$_\odot$ model
is shown in the upper panel of Figure \ref{fig:f19_examples} where one
can see that the thermonuclear production at the base of the He~shell
is confined to a relatively narrow region. The thermonuclear production of  $^{19}$F requires two 
components. First the presence of $^{18}$O that is a produced via
 $^{14}$N$(\alpha,\gamma)^{18}$F$(\beta^+ \nu_e)^{18}$O.
$^{14}$N results from the CNO cycle and thus the region for suitable for the thermonuclear
production of  $^{19}$F is sensitive to the physics of Hydrogen burning.
Secondly, the peak temperature reached in this region needs to be 
in the range of $0.4-0.5$ GK.

Assuming that internal energy
after shock passage is dominated by radiation one can relate the
explosion energy $E_{\text{expl}}$ and the peak temperature
$T_{\text{peak}}$ at a given radius $r$ can
as~\citep{Woosley.Heger.Weaver:2002}:

\begin{equation}
\label{eq:Tpeak}
T_{\text{peak}}=2.4 \left(\frac{E_{\text{expl}}}{10^{51}\
\text{erg}}\right)^{1/4} \left( \frac{r}{10^9\ \text{cm}}
\right)^{-3/4} \text{ GK}. 
\end{equation}
This illustrates that whether the optimal temperature conditions 
for $^{19}$F production are reached for a given progenitor abundance profile 
is very sensitive to the
radial position of the compositional shell interfaces and also mildly sensitive to the
explosion energy. The optimal temperature itself is determined by
thermonuclear reaction rates and recent updates on the proton capture
rates~\citep{Iliadis.Longland.ea:2010} have
a significant impact on the production of $^{19}$F.
Compared to calculations with reaction rates based on \citet{Caughlan.Fowler:1988} and \citet{NACR:1999} 
the updated reaction rates have increased the total
yield of $^{19}$F by 15\% without neutrinos and by 20\% with the
high neutrino energies for the
15~M$_\odot$ model. 

Trends of the  $^{19}$F production with respect to the progenitor mass can be related to
these sensitivities of the thermonuclear production.
With increasing initial stellar mass the $^{19}$F production factor
without neutrinos tends to decrease because of a larger production of
$^{16}$O even though the yield of $^{19}$F itself also increases
substantially as can be seen in Figure \ref{fig:stable_yields}.  The
$\nu$ process has also a larger impact because the mass contained in
the O/Ne layer increases while the thermonuclear production is
increasingly suppressed. 
This is illustrated with two examples in Figure~\ref{fig:f19_examples}
where the mass fraction of $^{19}$F for the 15 M$_\odot$ and 28
M$_\odot$ models are shown, representative for lower and upper end of
mass range considered here. For the 28~M$_\odot$ model the
contribution from the $\nu$ process in the O/Ne layer is the most
prominent effect of the explosion while the peak of thermonuclear
production at the inner He-shell in the 15 M$_\odot$ model gives an
important contribution.
For stars more massive than 17 M$_\odot$, the $\nu$ process can boost the $^{19}$F production by
factors of up to 1.5-2 and 3-4 for low and high energies respectively.

In contrast to the sensitivity of the thermonuclear production
mechanism to temperature and composition the $\nu$ process is mainly
sensitive to the distribution of $^{20}$Ne in the stellar model as
well as the cross sections for neutrino induced reactions on
$^{20}$Ne, which is now based on measured Gamow-Teller
strength~\citep{Anderson.Tamimi.ea:1991,Heger.Kolbe.ea:2005}. 
The remaining uncertainties in understanding the origin of  $^{19}$F are therefore
due to the stellar modeling and the thermonuclear reaction rates as well as the 
contribution from other astrophysical scenarios.

\subsection{Long-lived $^{138}$\textup{La} and nature's rarest element
  $^{180}$\textup{Ta}}
\label{sec:lata}

\begin{figure}[htb]
  \includegraphics[width=\linewidth]{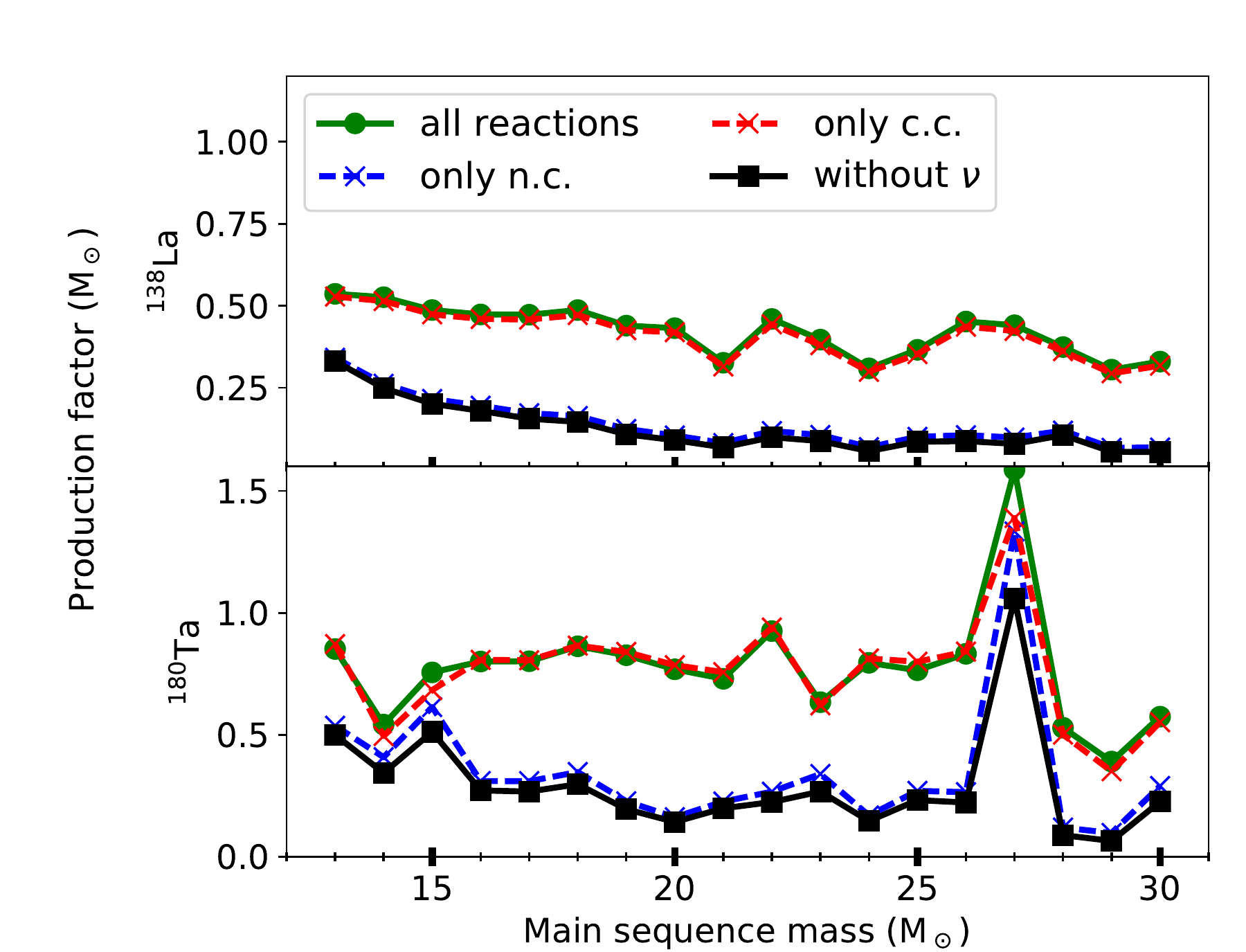}
  \caption{Production factors relative to $^{16}$O for $^{138}$La and
    $^{180}$Ta for the new set of 
  low neutrino energies.  Results including only neutral current
  (n.c.)
  and only charged current (c.c.) are also shown. Only for the
  14~M$_\odot$, 15~M$_\odot$ and 
  27~M$_\odot$ models a significant contribution from neutral current
  reactions 
  to the production of $^{180}$Ta appears while $^{138}$La is
  dominated by the c.c. reactions  
  for all the models. 
  The values for $^{180}$Ta correspond to the assumption that $35$\%
  of the yield   survive in the long-lived isomeric state. 
  \label{fig:lata_yields_total}}
\end{figure}

The production of the isotopes $^{138}$La and $^{180}$Ta is of
particular interest because they both are present in the solar system
but their production mechanism is not yet fully understood.
In contrast to the nuclei discussed above which are strongly affected by neutral current reactions, the $\nu$ process affects $^{138}$La and
$^{180}$Ta almost exclusively via $\nu_e$ captures. 
Therefore, those
two isotopes are the most promising species to infer $\nu_e$
properties via $\nu$ process nucleosynthesis. The cross sections for
$^{138}$Ba$(\nu_e,e^-)^{138}$La and $^{180}$Hf$(\nu_e,e^-)^{180}$Ta
are well constrained based on experimentally measured transition
strengths~\citep{Byelikov.Adachi.ea:2007}.

s-process nucleosynthesis
calculations~\citep{Belic.Arlandini.ea:2002,Kaeppeler.Arlandini.ea:2004}
have shown that around 80\% of the solar $^{180}$Ta can be produced in
AGB stars mostly via decays of excited states of $^{179}$Hf and
$^{180}$Hf.  However, \citet{Heger.Kolbe.ea:2005} have also found an
overproduction of $^{180}$Ta due to the $\nu$ process in core collapse
supernovae.

\begin{figure}
\includegraphics[width=\linewidth]{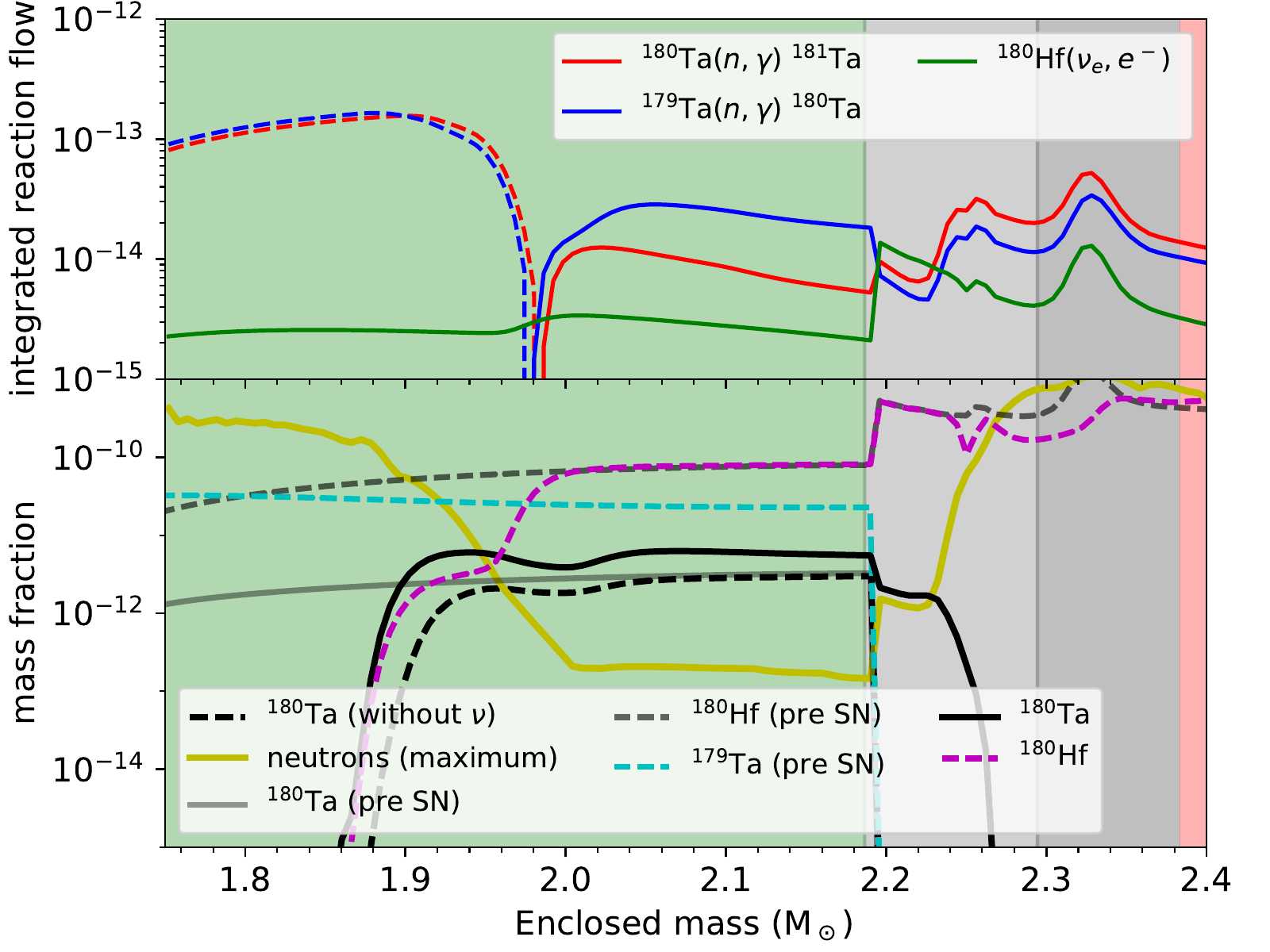}
\caption{Time integrated reaction flows (top panel) and mass fraction
  profile of $^{180}$Ta and relevant nuclei (bottom panel) for the 15
  M$_\odot$ model. If the forward $(n,\gamma)$ is dominating flows are
  shown as solid lines while dashed lines indicate that the inverse
  process $(\gamma,n)$ dominates. The neutrino induced reaction flow
  through $^{180}$Hf$(\nu_e,e^-)^{180}$Ta is also shown and only
  dominates in a very narrow region. The background colors indicate
  the different compositional layers as in Figure \ref{fig:LiB_s15}. }
\label{fig:ta180_reaction_flow}
\end{figure}

\begin{figure}[htb]
\includegraphics[width=\linewidth]{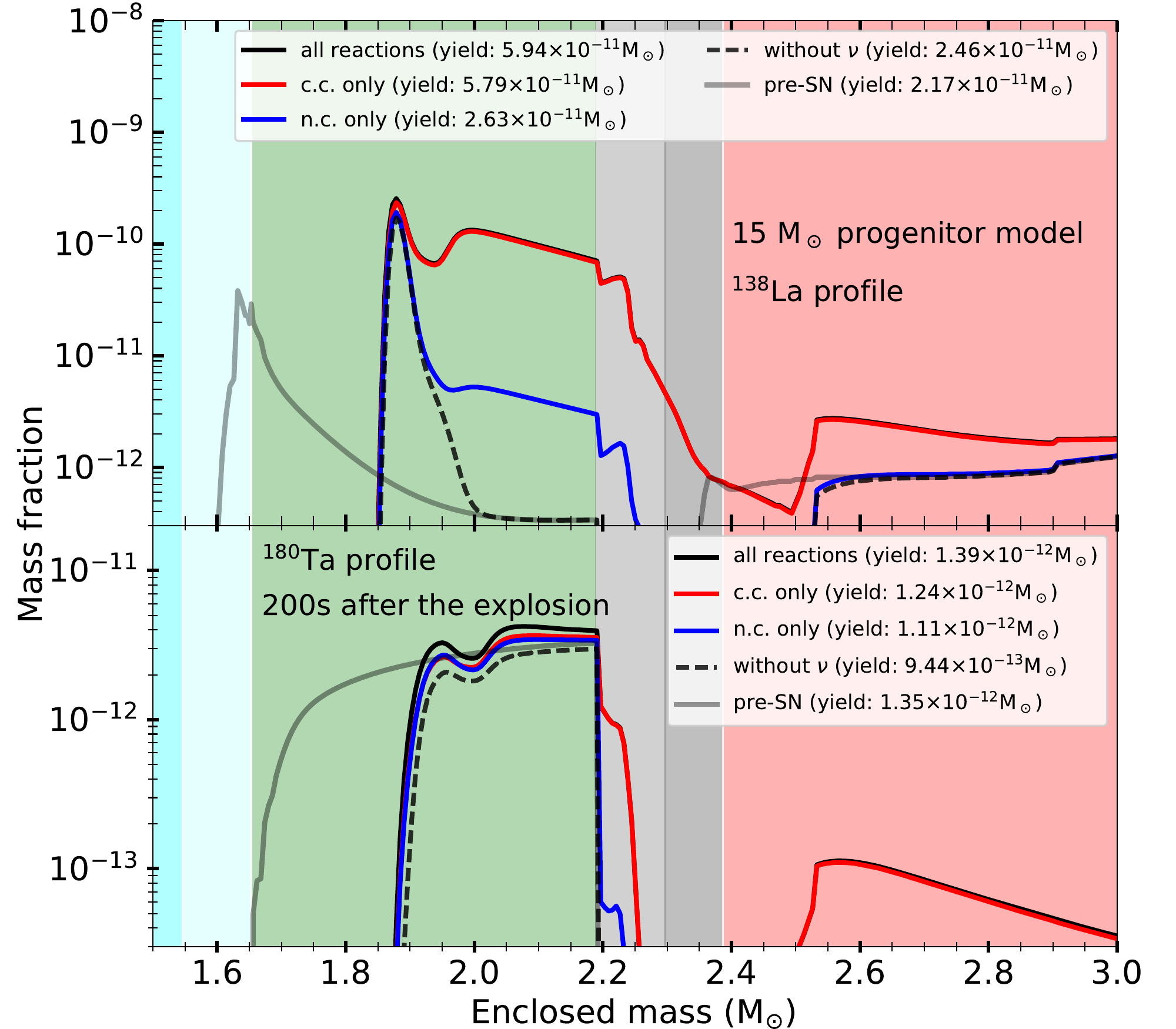}
\caption{Mass fraction profiles of $^{138}$La and $^{180}$Ta for the
  15 M$_\odot$ progenitor model. The background colors indicate the
  compositional zones as in \ref{fig:LiB_s15}.  Also shown are results
  with only neutral and charged current reactions.  $^{138}$La and
  $^{180}$Ta are both affected primarily by the charged current
  reactions with electron neutrinos. \label{fig:lata_yields}}
\end{figure}

Understanding the origin of $^{180}$Ta is further complicated by the fact that
the $J^\pi = 1^+ $ ground state $^{180}$Ta decays by electron capture and $\beta^-$ with a half-life of 8.15 h and only its isomeric $9^-$ state at an excitation energy of $75$ keV is very long lived.
Due to its high spin, the isomeric state is effectively decoupled from the ground state at low temperatures.
We do not treat $^{180}$Ta and its meta-stable isomeric
state $^{180}$Ta$^m$ as separate species in our network.
A significant fraction of the $^{180}$Ta the ground state has already decayed 
at the end of our nucleosynthesis calculations at $2.5 \times 10^4$~s. 
Therefore, we take the $^{180}$Ta abundance at $200$~s after the start of the 
calculation when most of the produced  $^{180}$Ta is still present and the 
thermal equilibrium between the ground state and the long-lived isomeric state has 
just frozen out. Following
the estimates derived by \citet{Mohr.Kaeppeler.Gallino:2007}
we assume that about 35\% of $^{180}$Ta
survives in the excited state.

Using the set of high neutrino energies, our
results for $^{138}$La and $^{180}$Ta are consistent with those of
presented by~\citet{Heger.Kolbe.ea:2005} and \citet{Byelikov.Adachi.ea:2007},
giving almost solar production of $^{138}$La and $^{180}$Ta. The $^{180}$Ta
production shown in table~\ref{tab:prodfac} and Figure~\ref{fig:lata_yields_total} 
are corrected for the fact that only the isomeric
state of $^{180}$Ta is long lived.

Figure \ref{fig:ta180_reaction_flow} illustrates the production of $^{180}$Ta
in the 15 M$_\odot$ model. The reaction flows have been estimated as integrated
instantaneous flows based on the abundances at the beginning of each time step
of the network calculation. At the base of the O/Ne shell, below 1.9 M$_\odot$
the peak temperature exceeds $2.6$ GK and the density reaches $5\times
10^5$~g~cm$^{-3}$.  Under these conditions not only the pre-supernova abundance
of $^{180}$Ta but also $^{180}$Hf are destroyed by photodissociation (see
Figure \ref{fig:lata_yields}).  Since $^{180}$Hf is the target for the $\nu_e$
captures, this marks the lower boundary of the production region.  The
prevalence of $(\gamma,n)$ over $(n,\gamma)$ is indicated by the
dashed lines in the upper panel of Figure
\ref{fig:ta180_reaction_flow}.  Further out, as the peak temperature
decreases below 2.3 GK, the $\gamma$ process leads to peak in the
production of $^{180}$Ta. The $\nu$ process increases the maximum
abundance but only operates after the shock has passed and the
material has cooled to below 2 GK.

For the particular case of the 15 M$_\odot$ model the pre SN abundance
of $^{179}$Ta is larger than the $^{180}$Ta abundance.  Therefore, not
only the direct charged current channel
$^{180}$Hf$(\nu_e,e^-)^{180}$Ta plays a role but also free neutrons
mainly from $^{16}$O$(\nu,\nu' n)$, $^{24}$Mg$(\nu,\nu' n)$ and
$^{20}$Ne$(\nu,\nu'n)$ increase the final yield of $^{180}$Ta.
Figure \ref{fig:lata_yields} shows that neutral current reactions
alone lead to a significant increase of the final yield even though
the contribution from $^{180}$Hf$(\nu_e,e^-)^{180}$Ta still dominates.

As the peak temperature is lower at a higher mass coordinate
the conditions for an effective production via the $\gamma$ process
are no longer reached and the time integrated reaction flows change
sign (see Figure~\ref{fig:ta180_reaction_flow}).
The competition between the
$^{179}$Ta$(n,\gamma)$ and $^{180}$Ta$(n,\gamma)$ is important for the
final abundance of $^{180}$Ta in this
region. \citet{Wisshak.Voss.ea:2001} have presented measurements for
the neutron capture cross section on $^{180}$Ta$^{m}$ and we use the
reaction rates from the KADoNiS v0.3 database~\citep{Dillmann.Szuecs.ea:2014} for both reactions.

At mass coordinate of 2.0 M$_\odot$
the main source for free neutrons changes from photodissociation to neutrino spallation. 
Without neutrinos, the initial $^{180}$Ta mass fraction remains unchanged in
this region.  Since this progenitor model provides a relatively high initial
abundance of $^{179}$Ta both processes, the direct charged current and the
neutral current providing additional free neutrons, are active. 

Only in a small region between 2.2 and 2.3 M$_\odot$ do the $\nu_e$ captures on
$^{180}$Hf dominate the production. Further out in the He shell, free neutrons
from $^{22}$Ne$(\alpha,n)$ destroy any $^{180}$Ta that is produced by
neutrinos, forming the upper boundary of the production region. The role of neutral current and charged current reactions depends
significantly on the progenitor structure.  In the range of progenitor
models studied here the 14, 15 M$_\odot$ and 27 M$_\odot$ models are
the only cases for which more than 10\% of the $\nu$ process
contribution to $^{180}$Ta results from neutral current neutrino
reactions due to the additional neutrons to be captured on $^{179}$Ta
because those progenitor models are already enriched in $^{179}$Ta and
$^{180}$Ta while at the same time depleted in $^{180}$Hf. The 27
M$_\odot$ stands out in particular because $^{180}$Ta is already
produced to full solar abundance before the explosion and without
neutrinos.  The $\nu$ process increases the production factor to 1.5
and 2.5 for low and high neutrino energies respectively. The pre-explosive production of $^{179,180}$Ta depends sensitively on the
temperatures reached during the last burning stages.  If the O/Ne
shell becomes hot enough, photodissociation can change the abundances
significantly. This shows that more detailed modeling of the pre
supernova phase is desirable to understand not only the explosion
mechanism as suggested by \citet{Suwa.Mueller:2016} but it might also
have a large effect on the synthesis of individual nuclear species.

As can be seen in Figure \ref{fig:lata_yields_total} the 14 and
15~M$_\odot$ stars also show a particularly low $\nu$ process contribution
to $^{180}$Ta and a relatively large $^{180}$Ta and low $^{180}$Hf
abundance before the explosion, possibly as a result of slightly
hotter burning conditions during the evolution.  (This is also
reflected in the $^{98}$Tc abundances shown in Figure
\ref{fig:radio_enhancement}.)  The reaction cross section for
$^{180}$Hf$(\nu_e,e^{-} n)^{179}$Ta is comparable to
$^{180}$Hf$(\nu_e,e^{-})^{180}$Ta for average neutrino energies around
10 MeV such that $^{179}$Ta can also be produced in-situ.  This
process contributes 10-20\% to the total $^{180}$Ta yield.
Averaged over the whole range of progenitors $^{180}$Ta is
underproduced with the new set of lower neutrino energies (see table
\ref{tab:prodfac}) ameliorating the tension with the contribution from
the $s$ process in AGB stars.

\begin{figure}
\includegraphics[width=\linewidth]{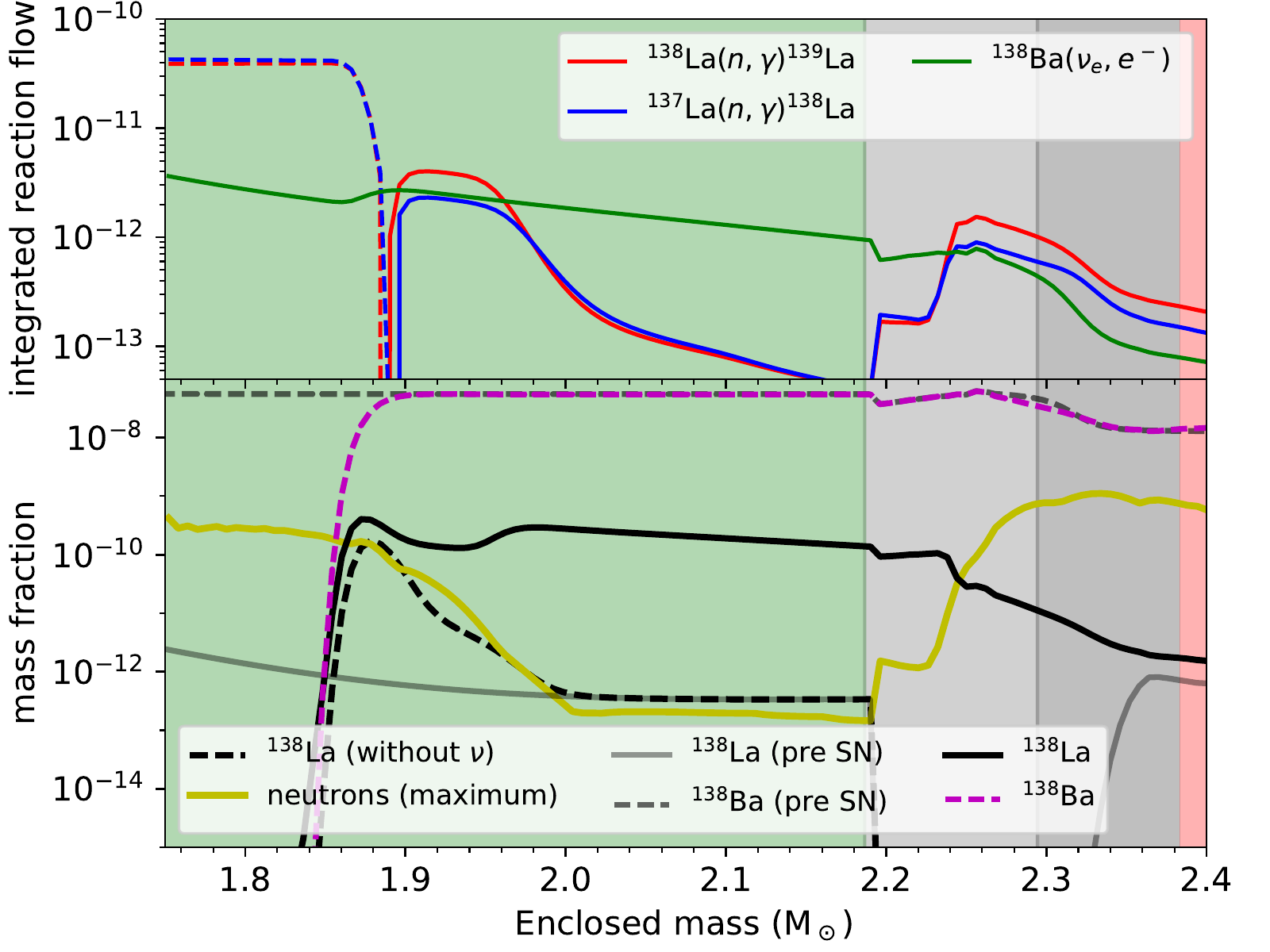}
\caption{Same as Figure \ref{fig:ta180_reaction_flow} but for
  $^{138}$La. There is also a peak in the production due to
  $(\gamma,n)$ reactions and in this case there is a region between
  2.0 and 2.3 M$_\odot$ where the $\nu$~process clearly is the
  dominant production channel.}
\label{fig:la138_reaction_flow}
\end{figure}

$^{138}$La is also a $p$~nucleus that is bypassed by the $s$~process
that moves along the chain of stable Barium isotopes. Figure
\ref{fig:la138_reaction_flow} shows the most important reaction flows
affecting $^{138}$La. Similar to the case of $^{180}$Ta, the production
of $^{138}$La at the base of the O/Ne layer is dominated by the
competition between $(n,\gamma)$ and $(\gamma,n)$ reactions, leading
to a peak of the production even without neutrinos due to
photodissociation of $^{139}$La at peak temperatures of around 2.4 GK.
As the peak temperature drops below 2 GK neutron captures dominate and
tend to move the material towards $^{139}$La. Without neutrinos, the
supply of $^{137}$La is very small. However,
$^{138}$Ba$(\nu_e,e^{-} n)$ can lead to a substantial production of
$^{137}$La because of the relatively large abundance of $^{138}$Ba.
The cross section for the reaction $^{138}$Ba$(\nu_e,e^- n)$ is based 
on the experimentally determined B(GT) strength \citet{Byelikov.Adachi.ea:2007}
in combination 
with branching ratios for particle emission from a statistical model as stated above.  

In contrast to the case of $^{180}$Ta free neutrons from neutral
current spallation reactions alone do not have a significant effect
because the relevant target nucleus $^{137}$La is for all progenitors
much less abundant than $^{138}$Ba and would need to be produced by
the charged current reaction first.  In our calculations about
10\% of the total yield of $^{138}$La result from neutron captures on
$^{137}$La. This contribution is sensitive to the ratio between the
$^{137}$La$(n,\gamma)$ and $^{138}$La$(n,\gamma)$ cross
sections. Therefore, we have taken the reaction rates by
\citet{Rauscher.Thielemann.ea:2000} in spite of recent experimental
constraints on the $^{138}$La$(n,\gamma)$ cross
section~\citep{Kheswa.Wiedeking.ea:2015}.

In Figure \ref{fig:la138_reaction_flow} one can see that further out
in the O/Ne shell where temperatures are too low to produce a
significant neutron density by photodissociation the direct neutrino
induced production is the dominating reaction flow, leading to an
extended region where the $^{138}$La mass fraction is almost
exclusively determined by $\nu_e$. Therefore, the production by the
$\nu$ process increases with the amount of mass in the O/Ne shell.

The upper panel of Figure \ref{fig:lata_yields_total} shows the
production factor for $^{138}$La over the range of progenitor models
discussed here with the set of our set of low
neutrino energies. Figure \ref{fig:lata_yields_total} also shows the results of calculations with
only neutral current and charged current reactions, illustrating that
the charged current channel clearly dominates over the whole range of
progenitors.  The most striking feature is a overly large production
of $^{138}$La for the 28 M$_\odot$ model.  This is due to a increased
production of Ba isotopes during the pre-supernova evolution. 
This progenitor is also enhanced in
weak $s$~process nuclei.
More massive progenitors contain more mass in the O/Ne layer and correspondingly give larger yields of $^{138}$La and
$^{180}$Ta. When looking at the production factors this increase of
the yield is compensated by an also increasing yield of $^{16}$O and a
the decreasing weight of more massive stars in the IMF.

$^{138}$La and $^{180}$Ta are mostly sensitive to electron flavor
neutrinos and since the production region in the O/Ne shell is closer
to the proto- neutron star they are also the most sensitive to
neutrino emission properties.  Therefore, those nuclei might
also might be affected the most by collective neutrino oscillations
as suggested by \citet{Wu.Qian.ea:2015}.
\subsection{Further effects on stable isotopes}
Recent studies dedicated to the  $\nu$~process have focused on individual nuclei
and have employed limited sets of neutrino-nucleus cross-sections expected to be
relevant for the nuclei of interest. In particular when such approaches focus on
a single progenitor model, the question whether there are additional effects 
in other scenarios or due to different reactions that have not been included always remains open.
With our complete set of cross sections we can survey the whole range of the reaction 
network at once and study for the first time the complete effect of $\nu$-reactions on the explosive
nucleosynthesis in supernovae for a whole range of progenitor models. 
The abundances of stable nuclei in the solar system are one of our
most accurately measured observables making processes that have an effect on those
nuclei particularly important.  
Unless major changes in the models for 
the progenitor composition or the supernova mechanism itself are found, we hope to have captured all possible processes and give in the following a summary of the changes of the yields of stable nuclei after decay, before we enter on the discussion of radioactive isotopes in \S\ref{sec:radio}.

Table \ref{tab:maxes} summarizes the maximum differences $\delta_{rel}=(Y_{\mathrm{no}\,\nu}-Y_{\nu})/Y_{\mathrm{no}\,\nu}$ in the integrated yields of stable and very long-lived ($T_{1/2}>10^{10}$ years) nuclei after decay that we find among all the progenitor models studied here. The table shows
that large effects that change abundances by a significant factor indeed only appear 
for nuclei that have been identified in previous studies. On the 10\% level we find
a few more nuclei that are affected. In most cases the 
maximum effect is found for the more massive progenitors. That is because the inner regions of more massive stars tend to be more compact, putting the relevant O/Ne layer closer to the PNS. Only the light isotopes  $^{7}$Li and $^{9}$Be that are produced at larger radii in the He-shell are maximally affected in at the low mass end of our progenitor range because here the He-shell is at smaller radii. 
Even though $^{9}$Be and $^{10}$B are listed in Table \ref{tab:maxes} their yields correspond to production factors of at most $4.5\times 10^{-2}$ and $7\times 10^{-2}$ respectively, too low to explain their solar abundances or solar ratios with respect to $^{7}$Li and $^{11}$B. However, models of GCR nucleosynthesis can account for those nuclei as stated in \S\ref{sec:light-elements-li}.
A modification of $^{17}$O mass fraction is found throughout the O/Ne shell and is mostly induced by neutron captures on abundant $^{16}$O where the 
neutrons are 
released by neutral-current spallation reactions. Locally the mass fraction of $^{17}$O is increased by several orders of magnitude.
However, the total yield is dominated by the abundance of  $^{17}$O in the He-shell left over from H-burning via the CNO cycle.
The modification of $^{33}$S occurs in the Si/O shell and it is affected by several reactions,
including the reaction sequence $^{34}$S$(\nu_e,e^-)^{34}$Cl$(\gamma,p)$ as well as $^{34}$S$(\nu,\nu' n)^{33}$S. Thus, the contributions of charged- and neutral-current reactions is about equal.
At the top of the O/Si shell  $^{35}$Cl is enhanced by $^{36}$Ar$(\nu,\nu' p)$ where  $^{36}$Ar is a results of the $\alpha$-rich freeze out. The production factor for  $^{35}$Cl on average around $0.5$, making the contribution from massive stars a relevant for the solar system inventory of  $^{35}$Cl.
The yield of $^{41}$K is modified mostly by $^{41}$Ca$(\bar{\nu}_e,e^+)$ and to a lesser extent by $^{42}$Ca$(\nu,\nu' p)$. Averaged over the range of progenitors the $\nu$~process increases the production factor for  $^{41}$K from $0.48$ to $0.52$ for the low neutrino energies but it reaches values of up to $1.8$ for the 20 M$_\odot$ progenitor for which the effect of neutrinos is negligible. 
$^{176}$Lu is affected by electron antineutrino capture on $^{176}$Hf which is inherited from the initial metallicity in the O/Ne shell, very similar to the cases of $^{138}$La and 
$^{180}$Ta but interestingly involving antineutrinos in this case. 
The IMF averaged production factor is however below $0.2$.  $^{176}$Lu can be explained with the main $s$-process in AGB-stars and subject of current experimental efforts \citep{Roig.Jandel.ea:2016}. The case of $^{176}$Lu is further complicated by a short-lived $1^-$ excited state at $122$~keV above the $7^-$ ground state that $\beta$ decays with a half-life of 3.7~h. The short lived isomer is likely to be populated thermally under supernova conditions and since we do not include it explicitly in our calculations we expect that our results overestimate the yield of $^{176}$Lu.
In contrast to that $^{113}$In is a p-nucleus that is also produced via the $\gamma$-process.
Neutrinos affect its yield by $\nu_{\!e}$ captures on $^{113}$Cd in the O/Ne shell where the mass fraction is increased by up to a factor of $500$ to values of up to $3\times 10^{-12}$ which is still a factor ten smaller than the abundance inherited from the initial metallicity.
This isotope is particularly interesting because \citet{Travaglio.Roepke.ea:2011} found it to be underproduced in type Ia supernovae. However, in our calculation
we also find a production factor of at most $0.32$ because the abundance in the O/Ne shell is still low compared to the solar abundance. The optical model potentials to describe the involved  $(\gamma,\alpha)$ reactions have recently been studied by \citet{Kiss.Mohr.ea:2013} where a good agreement of the total cross-sections with the theoretical calculations was found.
We find that the final integrated yield of $^{59}$Co is reduced by 11\% in the 15 M$_\odot$ model. This is due to $^{59}$Ni$(\nu_e,e^-n)^{58}$Cu which reduces the abundance of the long-lived $^{59}$Ni with a half-live of $7.6\times 10^4$ yr that finally decays to $^{59}$Co.
The modification of the  $^{57}$Fe results mostly from the charged-current reaction $^{58}$Ni$(\bar{\nu}_e,e^+ p)$ and also involves $^{58}$Co$(\nu,\nu' p)$. The 12\% increase in the yield of $^{54}$Cr for the 23 M$_\odot$ reflects the production of $^{54}$Mn by $\bar{\nu}_{e}$ capture on $^{54}$Fe that reaches a mass fraction of $5\times 10^{-2}$ the O/Si shell. 
We see that in the O/Si shell close to the mass~cut reactions on the Fe-peak elements induce some changes on 
the ejecta composition on the order of few percent.
However, since our piston model is not expected to give a very good description of these innermost regions \citep{Young.Fryer:2007} that 
are sensitive to the imposed mass cut and potential fallback of material, further studies with
self-consistent explosion models are needed to verify the significance of these effects.
\citet{Heger.Kolbe.ea:2005} have suggested reactions that could modify the yields of $^{51}$V, $^{55}$Mn, $^{78}$Kr, $^{138}$Ce and $^{196}$Hg. Our calculations include all the reactions suggested by \citet{Heger.Kolbe.ea:2005} and we find that the yields of these nuclei are increased by 5-9\%.
The effects on the $p$-nuclei  $^{113}$In,  $^{137}$La and  $^{180}$Ta shown here also illustrate that it is necessary to include neutrino-induced reactions for 
quantitative predictions of $\gamma$-process nucleosynthesis.

\begin{table}[htb]
\centering
\caption{Maximum values of the relative difference
$\delta_{rel}^{max}=(Y_{\nu}-Y_{\mathrm{no}\, \nu})/Y_{\mathrm{no}\,\nu}$ that are larger than 10\% from all progenitor models
considered here for the set of low neutrino energies. Also shown are the yields
in M$_\odot$ for the calculations with and without including neutrino
reactions. The last column gives the mass of the progenitor for which the
maximum value is found  $M_*^{max}$.}
 \begin{ruledtabular}
 \begin{tabular}{lcccc}
 Nucleus & $\delta_{rel}^{max}$(\%) & $ Y_{\mathrm{no}\,\nu} $(M$_\odot$)   & $Y_\nu$(M$_\odot$) & $M_*^{max}$ (M$_\odot$) \\ \hline
 $^{7}$Li    & $  2,250$ & $  1.69\times 10^{-9}$ &  $  3.96\times 10^{-8}$ &  15  \\  
$^{9}$Be    & $   25$ & $  7.19\times 10^{-11}$ &  $  8.97\times 10^{-11}$ &  18  \\  
$^{10}$B    & $   34$ & $  1.79\times 10^{-9}$ &  $  2.40\times 10^{-9}$ &  25  \\  
$^{11}$B    & $  8,208$ & $  5.78\times 10^{-9}$ &  $  4.80\times 10^{-7}$ &  25  \\  
$^{15}$N    & $   188$ & $  2.58\times 10^{-5}$ &  $  7.45\times 10^{-5}$ &  30  \\  
$^{17}$O    & $   16$ & $  5.98\times 10^{-5}$ &  $  6.91\times 10^{-5}$ &  30  \\  
$^{19}$F    & $   88$ & $  6.92\times 10^{-6}$ &  $  1.30\times 10^{-5}$ &  20  \\  
$^{33}$S    & $   14$ & $  3.75\times 10^{-4}$ &  $  4.26\times 10^{-4}$ &  19  \\  
$^{35}$Cl   & $   11$ & $  3.91\times 10^{-4}$ &  $  4.35\times 10^{-4}$ &  25  \\   
$^{41}$K    & $   22$ & $  2.82\times 10^{-5}$ &  $  3.44\times 10^{-5}$ &  19  \\  
$^{54}$Cr   & $   12$ & $  3.07\times 10^{-5}$ &  $  3.43\times 10^{-5}$ &  23  \\  
$^{57}$Fe   & $   13$ & $  4.66\times 10^{-3}$ &  $  5.25\times 10^{-3}$ &  25  \\  
$^{59}$Co   & $   -12$ & $  8.42\times 10^{-4}$ &  $  7.39\times 10^{-4}$ &  17 \\  
$^{78}$Kr   & $   10$ & $  3.63\times 10^{-8}$ &  $  3.26\times 10^{-8}$ &  23 \\   
$^{113}$In  & $   19$ & $  5.44\times 10^{-10}$ &  $  6.48\times 10^{-10}$ &  27 \\  
$^{138}$La  & $   511$ & $  5.35\times 10^{-11}$ &  $  3.27\times 10^{-10}$ &  30 \\  
$^{176}$Lu  & $   14$ & $  1.79\times 10^{-10}$ &  $  2.04\times 10^{-10}$ &  30 \\  
$^{180}$Ta  & $   501$ & $  4.93\times 10^{-13}$ &  $  2.96\times 10^{-12}$ &  29 \\
 \end{tabular}
 \end{ruledtabular}
\label{tab:maxes}
\end{table}

Unless significant changes in the progenitor composition or the neutrino properties are found, we can thus exclude and further effects on stable nuclei due to the $\nu$~process on supernova nucleosynthesis for stars in the mass range 13-30 $M_\odot$ at solar metallicity. 
\section{Radioactive nuclei}
\label{sec:radio}

\subsection{Overview}
\label{sec:radio_overview}
\begin{figure}[htb]
  \includegraphics[width=\linewidth]{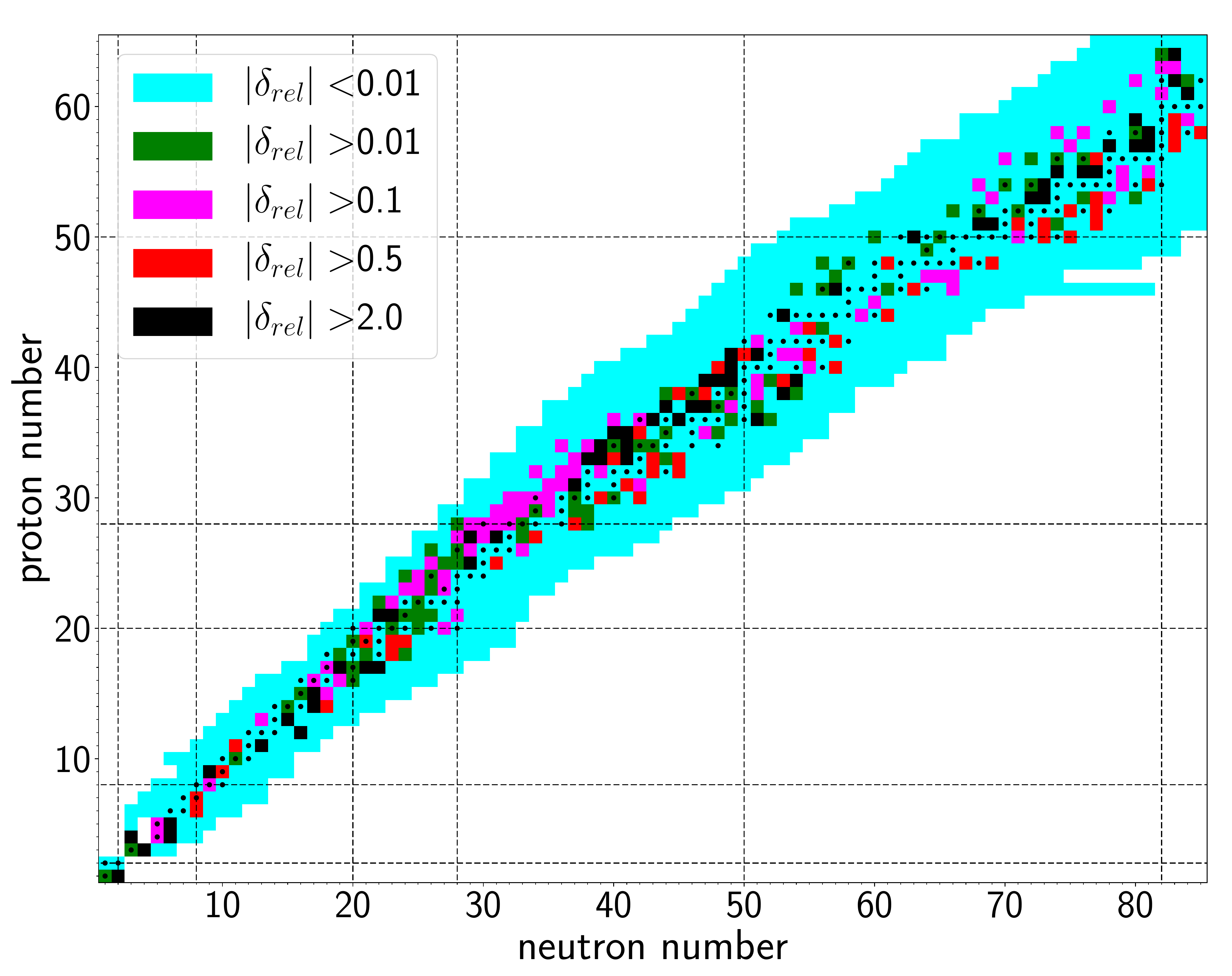}
  \caption{Maximum values among all the progenitors studied here for the absolute of the relative change $\delta_{rel}$ as defined in Figure \ref{fig:al26na22}
    with the new set of low neutrino
    energies at 2.5 $\times 10^{4}$ s after core bounce when the very short lived nuclei have already decayed. Only nuclei with
    a mass fraction larger than $10^{-12}$ are included.  \label{fig:radio_effects}}
\end{figure}

In addition to the important effect on the yields of stable isotopes we will see in the following that the $\nu$~process affects the 
production of many radioactive nuclei.
Mostly, the modification of the abundances of radioactive nuclei does not result in noticeable changes of the yields of stable nuclei
after the radioactive isotopes have decayed. 
However, for some species
the decay is accompanied by the emission of characteristic $\gamma$-rays 
and others leave traces in the composition of pre-solar grains. The most interesting cases are discussed in
detail in \S\ref{sec:clnbtc} and \S\ref{sec:alna} but before we give an overview of the effects on radioactive nuclei, focusing on the  15 M$_\odot$ progenitor model as a representative example for 
the mass range we have explored.

We find that a large range of radioactive species are substantially affected by the
$\nu$~process. 
Figure~\ref{fig:radio_effects} provides an overview of the relative
effects of the $\nu$~process for
the whole range of nuclei included in our calculations at 7 hours after the explosion when very 
short lived nuclei have already decayed. The relative differences $\delta_{rel}$ shown there, are the
maximum values we find for the whole range of progenitor models we have looked at. The largest
effects appear close to stability where seed nuclei with a large abundances relative
to their neighbors are present. 
Many nuclei are affected on the 10\% level and a few show differences exceeding
50\% or a factor 2. Below the Iron group the $\nu$ process
mostly increases the production of isolated rare stable and long-lived
nuclei discussed in the previous sections. Spallation reactions on the most abundant nuclei like $^{16}$O
$^{20}$Ne and $^{24}$Mg do not change the abundances of the targets noticeably but
the neighboring nuclei get produced and they
provide light particles that affect other nuclei. 
Additional neutrons are
mostly captured by heavier nuclei, leading to increased abundances on the
neutron rich side for $A>100$ where the seed nuclei are inherited from the initial solar 
metallicity. Since the $\gamma$ process also operates on those seed nuclei, 
the abundances on the proton-rich side are also modified slightly.
Around the Iron group many long-lived nuclei exist and they are mainly 
produced in the Si-shell close to the PNS where the neutrino fluxes are largest. After freeze out from NSE, neutrino interactions
reshuffle the abundances of the Fe-peak with differences at the 10\% level, leading for example to
the modification of the $^{59}$Co yields discussed previously. 
In the region of $A=60-90$ we can see a significant modification of the abundances, both on the
neutron- and proton-rich side of stability. This is due to the weak $s-process$ nuclei and 
the operation of the $\gamma$-process that already leads to the production of radioactive nuclei in that region which are 
then modified further by neutrino interactions. Radioactive, neutron-deficient isotopes of As, Br, Kr, Sr, Y and Zr are particularly enhanced with 
mass fraction typically between $10^{-12}$ to $10^{-10}$. For progenitors with a weaker $\gamma$ process free neutrons from spallation reactions are captured on the most abundant $s$-process
nuclei and lead to increased abundances on the neutron-rich side in the same mass region.

The production of radioisotopes by the $\nu$~process has hitherto
received only limited attention in the literature which has mostly
focused on the five isotopes discussed above. In the following we will
discuss the overall effect of the $\nu$ process on the production of
radioactive nuclei, in particular focusing on those that are relevant for
observations.

\begin{table}[htb]
  \caption{Impact of the $\nu$ process on the yields of radioactive
    isotopes at the end of our calculation at $2.5\times10^4$~s. At this time
    the very short-lived nuclei have already decayed and mostly species that are
    potentially interesting for observations remain.
    Shown are the yields in units of M$_\odot$ averaged
    with an IMF as above, obtained without neutrino, with our choice
    of neutrino temperatures (``Low energies''), and with the choice
    of~\citet{Heger.Kolbe.ea:2005} (``High
    energies'').\label{tab:prodall}}
  \begin{ruledtabular}
    \begin{tabular}{llcccc}
      Nucleus& $T_{1/2}$ &no $\nu$ & Low energies\footnote{$T_{\nu_e}=
                          2.8$~MeV,
                          $T_{\bar{\nu}_e}=T_{\nu_{\mu,\tau}}=
                          4.0$~MeV}& High energies\footnote{$T_{\nu_e}
                                     = 4.0$~MeV, $T_{\bar{\nu}_e} =
                                     5.0$~MeV,
                                     $T_{\nu_{\mu,\tau}}=6.0$~MeV}\\       \hline  
      $^{22}$Na& 2.61 yr & 1.89$\times 10^{-6}$  &  2.42$\times 10^{-6}$  &  3.01$\times 10^{-6}$  \\
      ${}^{26}\!$Al& 0.72 Myr& 3.88$\times 10^{-5}$  &  4.19$\times 10^{-5}$  &  4.74$\times 10^{-5}$  \\
      $^{36}$Cl& 0.30 Myr& 2.89$\times 10^{-6}$  & 4.19$\times 10^{-6}$  &  5.01$\times 10^{-6}$  \\
      $^{44}$Ti& 59.1 yr & 3.68$\times 10^{-5}$  &  5.05$\times 10^{-5}$  &  5.17$\times 10^{-5}$  \\
      $^{60}$Fe& 2.6 Myr & 7.20$\times 10^{-5}$  &  7.21$\times 10^{-5}$  &  7.23$\times 10^{-5}$  \\
      ${}^{72}\!$As&26.0 h   & 2.38$\times 10^{-10}$  &  3.01$\times 10^{-9}$  &  7.48$\times 10^{-9}$  \\
      $^{84}$Rb&32.8 d   & 3.97$\times 10^{-10}$  &  2.87$\times 10^{-9}$  &  5.50$\times 10^{-9}$  \\
      $^{88}$Y &106.6 d  & 4.14$\times 10^{-10}$  &  1.27$\times 10^{-9}$  &  2.49$\times 10^{-9}$  \\
      $^{92}$Nb&34.7 Myr & 3.30$\times 10^{-11}$  &  7.38$\times 10^{-11}$  &  1.30$\times 10^{-10}$  \\
      $^{98}$Tc&4.2 Myr  & 2.57$\times 10^{-11}$  &  2.98$\times 10^{-11}$  &  3.61$\times 10^{-11}$  \\
    \end{tabular}
  \end{ruledtabular}
\end{table}

Table~\ref{tab:prodall} lists the IMF averaged nucleosynthesis yields
for a range of radioactive nuclei that are still present at around 7h after the explosion, including $^{32}$P,
$^{72}$As, $^{84}$Rb, $^{88}$Y. Their yields are increased by factors
between 2 and 10 with the realistic neutrino energies and the
production of $^{72}$As would be increased by almost two orders of
magnitude with the choice of high neutrino energies.  The
typical yields for $^{72}$As, $^{84}$Rb, and $^{88}$Y are
$10^{-9}$~M$_\odot$, which may allow for the observation of the
gamma-ray decay lines only with very high precision observations.
Further complicating the detection, their lifetimes are of the order
of a 100 days or shorter, putting their decay signal in competition
with $^{56}$Ni and its daughter $^{56}$Co which by far dominates the
early lightcurve and therefore outshines the signature of the $\nu$
process. However, this shows that the $\nu$ process can affect a large range of
radioactive nuclei among which we can look for a suitable candidate 
to provide an observable signature of supernova neutrinos.

\subsection{Short-lived radionuclides $^{36}$\textup{Cl}, $^{92}$\textup{Nb} and $^{98}$\textup{Tc} in the late input scenario }
\label{sec:clnbtc}
Isotopic ratios mostly derived from mass spectroscopy of
grains of meteoritic material
have proven to be an invaluable source of information on
stellar nucleosynthesis \citep{Zinner:1998,McKeegan.Davis:2003,Dauphas.Chaussidon:2011}.
While $^{138}$La and $^{180}$Ta are measurable as part of the current
composition of the solar system, indications for the presence of
now extinct radioactive nuclei, such as $^{36}$Cl and $^{92}$Nb have been found in
\citep{Schoenbaechler.Rehkaemper.ea:2002,Lin.Guan.ea:2005,Jacobsen.Matzel.ea:2009}
primitive meteorites that are assumed to have conserved the
composition of the material from which the solar system has
formed. 
For $^{98}$Tc a positive detection is still missing, but upper limits 
are given by \citet{Dauphas.Marty.ea:2002} and \citet{Becker.Walker:2003}.
In the following we discuss the production of $^{92}$Nb,
$^{98}$Tc and $^{36}$Cl in detail and compare the observed abundance ratios
from primitive meteorites.

\begin{figure}[htb]
  \includegraphics[width=\linewidth]{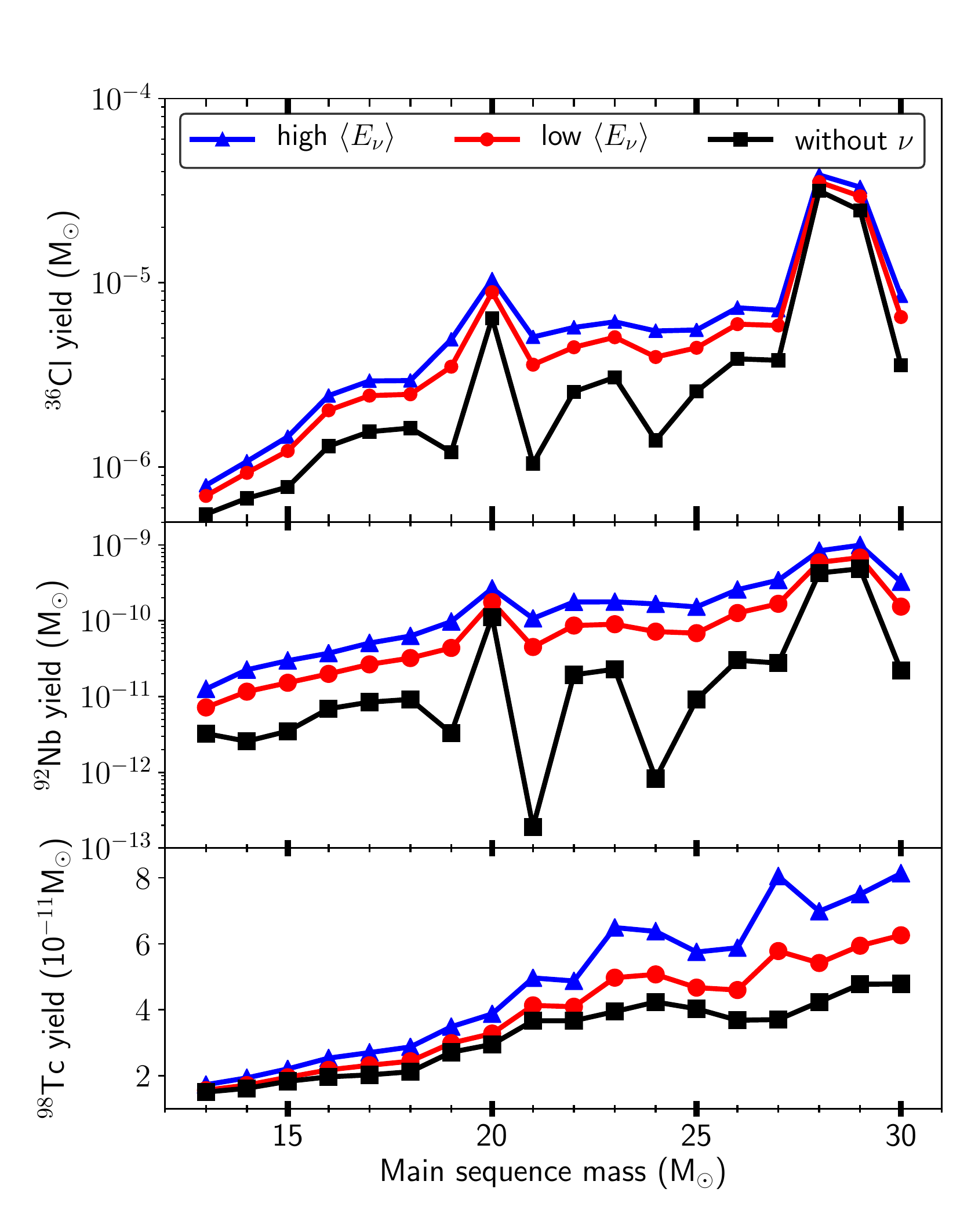}
  \caption{Yields of $^{36}$Cl, $^{92}$Nb and $^{98}$Tc for different sets of
  neutrino energies and without the $\nu$ process over the range of progenitors
  considered here. Note that the upper two panels are on a logarithmic scale
  white the $^{98}$Tc mass fraction is on a linear scale because it shows a
  much smaller variation.}\label{fig:radio_enhancement}
\end{figure}

\begin{figure}[htb]
\includegraphics[width=\linewidth]{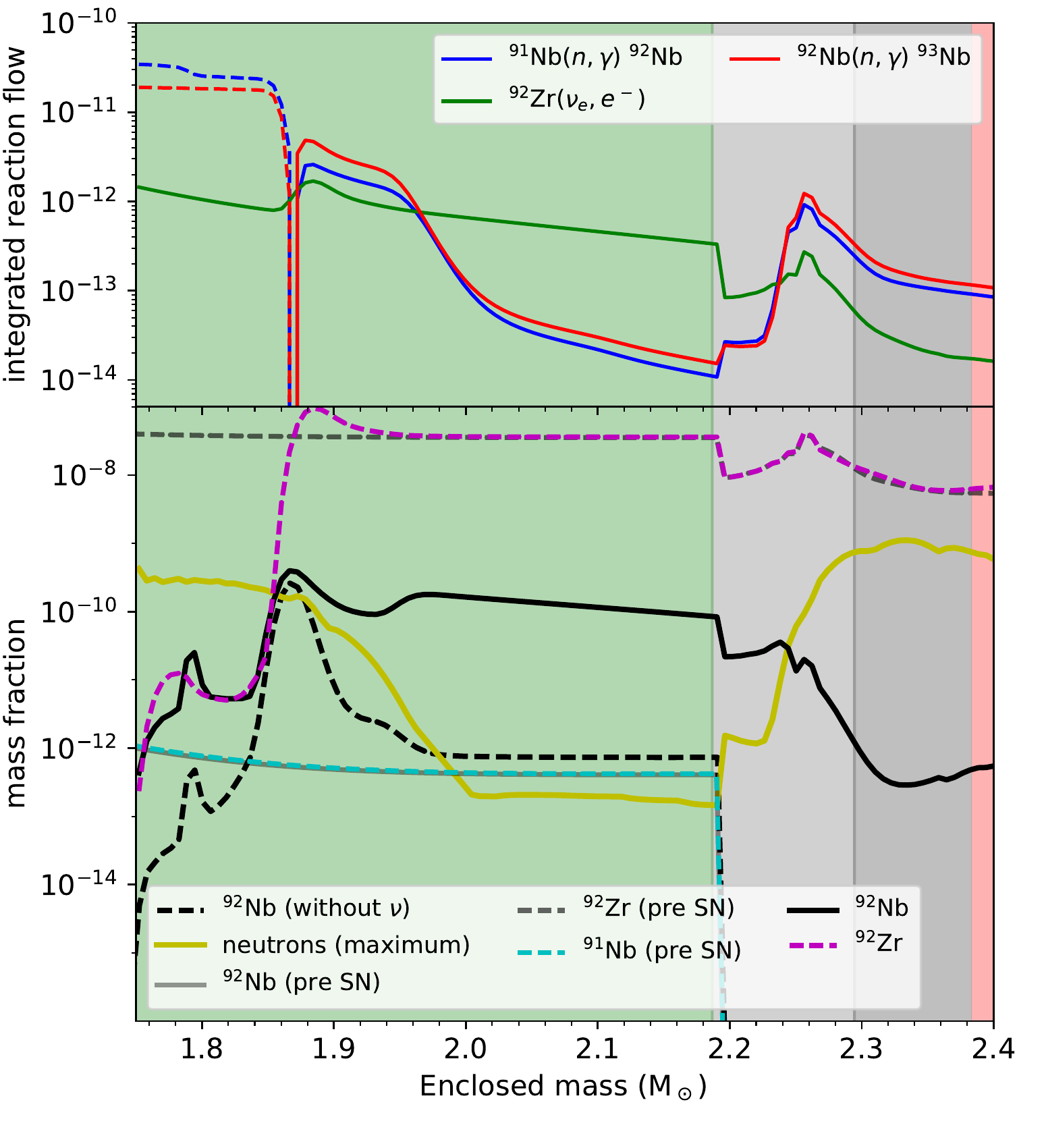}
\caption{Same as Figure \ref{fig:la138_reaction_flow} for $^{92}$Nb. The top panel shows time integrated
reaction flows and the lower panel gives an overview over the mass fractions of the involved nuclei. While the 
production of  $^{92}$Nb proceeds via $(n,\gamma$)-$(\gamma,n)$ 
reactions in the deeper, hotter part of the O/Ne shell, the $nu$~process
leads to a moderate production throughout the outer part of the O/Ne shell.}
\label{fig:nb92_reaction_flows}
\end{figure}

Our calculations show (see Table \ref{tab:prodall}), that neutrino
interactions increase the average yield of $^{36}$Cl by factors of
1.76 and 2.1 for low and high neutrino energies respectively and
$^{92}$Nb by a factors of $2.1$ and $3.7$. This relatively large
increase that does not vary a lot with the neutrino energies indicates
that the $\nu$ process provides a dominant and independent
contribution. \\
Even though the total yield of $^{98}$Tc is increased by less than 20\% 
locally the mass fraction of $^{98}$Tc in the O/Ne shell is
typically increased by one to two orders of magnitude because 
the total yield is dominated by the
pre-supernova content of $^{98}$Tc in the He-shell.
The enhancement is mostly due to
$^{98}$Mo$(\nu_e,e^-)$. Despite the higher average energy of electron
antineutrinos $^{98}$Ru$(\bar{\nu}_e,e^+)$ is negligible because
$^{98}$Ru is very rare with mass fractions typically around $10^{-16}$,
while $^{98}$Mo is much more abundant with mass fractions around
$5\times 10^{-9}$.  A particularly large amount of $^{98}$Mo is
present in the 27 M$_\odot$ which is in enriched in $\gamma$~process
nuclei as discussed for the case $^{180}$Ta in \S\ref{sec:lata}.
For most progenitor models there is also a direct contribution of the
$\gamma$~process during the shock heating to $^{98}$Tc which is less
than $10$\% of the $\nu$~process contribution. 

Figure \ref{fig:radio_enhancement} shows the yields of $^{36}$Cl,
$^{92}$Nb and $^{98}$Tc for the stellar models we have studied.
$^{36}$Cl and $^{92}$Nb exhibit very similar systematics with respect
to the progenitor mass because they are both very sensitive to the
composition and temperature at the inner edge of the O/Ne
shell. $^{36}$Cl and $^{92}$Nb are the two species with the deepest
$\nu$~process production region.  While their yields including the
$\nu$~process are relatively smooth with respect to the initial
progenitor mass, large variations can be seen in the calculations
without neutrinos.  As a result, also the relative enhancement ranges
from factors of 2-5 for most progenitors up to a factor of $600$ for
the 24 M$_\odot$ model which yields particularly little $^{92}$Nb
without the $\nu$~process.  The production mechanisms for $^{92}$Nb
and $^{98}$Tc in the $\nu$~process are very similar to the production
of $^{138}$La and $^{180}$Ta.  The upper panel of Figure
\ref{fig:nb92_reaction_flows} shows the dominating reaction flows
relevant for the synthesis of $^{92}$Nb in the 15 M$_\odot$ progenitor
model. At the bottom of the O/Ne shell, photodissociation and neutron
captures compete and in an optimal temperature range the $^{92}$Nb
mass fraction forms a peak even without neutrinos. The $\nu$~process
contributes evenly through the whole region which contains $^{92}$Zr
with a mass fraction of around $3\times 10^{-8}$.

The $^{36}$Cl yield without neutrinos results mostly from neutron
captures on $^{35}$Cl at lower to mid O/Ne shell. $^{35}$Cl is present
in the progenitor but is also produced by the shock heating. Providing
additional neutrons, neutrino neutral current spallation reactions
have a minor effect on the yield while
$^{36}$Ar$(\bar{\nu}_e,e^-)^{36}$Cl is the most important contribution
of the $\nu$~process for all of the progenitor models. During
O-burning $^{36}$Ar is copiously produced reaching mass fractions of
the order $10^{-2}$ in the final Si-shell. This $^{36}$Ar-rich
is exposed to temperatures exceeding $3$~GK and does not contribute to the $\nu$~process. In
the $15$M$_\odot$ model $^{36}$Ar is efficiently produced by
$^{35}$Cl$(p,\gamma)$ and $^{32}$S$(\alpha,\gamma)$ at the bottom of
the O/Ne shell where the peak temperature reaches up to 2.5 GK. While
for most other nuclei that are affected by the $\nu$~process, the
parent nucleus is already present in the progenitor, $^{36}$Ar and
also $^{35}$Cl first need to be produced by the shock heating. Hence,
$^{36}$Cl is also particularly sensitive to the shock propagation and
the explosion energy.

For the progenitors more massive than $15$ M$_\odot$ the production region 
moves to smaller radii, into the upper part of the Si-shell which also 
contains substantial amounts of Oxygen. The $20$ M$_\odot$ model stands out 
with a rather large pre-supernova production. In this model the 
$\nu$~process contribution is strongest in the Si/O-O/Ne transition 
region which consists of Si and Ne in roughly equal amounts. 

\citet{Cheoun.Ha.ea:2012} and \citet{Hayakawa.Nakamura.ea:2013} have
discussed the $\nu$ process in supernovae as a production site for the
radioactive isotopes $^{92}$Nb and $^{98}$Tc. In particular $^{92}$Nb
is interesting as a potential chronometer. \citet{Mohr:2016} has analyzed
the impact of an isomeric state of $^{92}$Nb at 135.5 keV on its
nucleosynthesis in an explosive environment and found that it does not
affect the production.  The survival might however be affected by a
reduced lifetime at low temperatures.
The yields of $^{92}$Nb and $^{98}$Tc might even be more enhanced by
contributions from the neutrino-driven
wind~\citep{Fuller.Meyer:1995,Hoffman.Woosley.ea:1996} which we do not include here.

Due to their long lifetimes and very low abundance  $^{92}$Nb and $^{98}$Tc are not 
suited for $\gamma$ ray astronomy and $^{36}$Cl decays mainly to the ground-state
of $^{36}$Ar without characteristic $\gamma$-rays.
Therefore, we need other observational constraints if we want to use the 
production of these nuclides to learn about supernova neutrinos.
Evidence for the presence of $^{92}$Nb \citep{Schoenbaechler.Rehkaemper.ea:2002} and $^{36}$Cl \citep{Jacobsen.Matzel.ea:2009} at the time 
the solar system has formed have been found in meteoritic grains (see also \citet{Wasserburg.Busso.ea:2006} for an overview).

\citet{Hayakawa.Nakamura.ea:2013} have estimated the contribution of
the $\nu$~process to the ISM inventory of $^{92}$Nb/$^{93}$Nb based on
11~M$_\odot$ supernova model. They conclude that the continuous uniform production is insufficient to
explain the isotopic ratio of $\approx 10^{-5}$ inferred from
primitive meteorites
\citep{Schoenbaechler.Rehkaemper.ea:2002}. 
While the estimate by \citet{Hayakawa.Nakamura.ea:2013} is based on a single progenitor model we
can use the IMF weighted average of the stellar models we have studies. We get an average ratio of $\langle ^{92}$Nb/$^{93}$Nb$\rangle=7.2 \times 10^{-4}$
without neutrinos. This reaches $1.5 \times 10^{-3}$ and
$2.6 \times 10^{-3}$ for low and high neutrino energies respectively.
Assuming a uniform production model and taking supernovae as the sole production site 
for both  $^{92}$Nb and $^{93}$Nb we estimate the ratio as \citep{Huss.Meyer.ea:2009}:
\begin{equation}
\label{eq:up}
\left(\frac{ \; X(^{92}\text{Nb})}{X(^{93}\text{Nb})} \right)_{UP} \approx 2 \langle ^{92}\text{Nb}/^{93}\text{Nb}\rangle \frac{
\tau_{^{92}\text{Nb}}}{\quad T}
\end{equation} 
Where $\tau_{^{92}\text{Nb}}=50.1$~Myr is the decay timescale of $^{92}$Nb. 
With a typical isolation time $T=10$ Gyr we get a maximum ratio of $1.3 \times 10^{-7}$ in agreement with 
\citet{Hayakawa.Nakamura.ea:2013} and still insufficient to explain the observed ratio.
\citet{Hayakawa.Nakamura.ea:2013} conclude that
a late injection event where the pre-solar material is polluted by
the ejecta from a nearby supernova is more likely. 
\citet{Banerjee.Qian.ea:2016} have recently consolidated this scenario using the short-lived radioactive $^{10}$Be produced by
the $\nu$~process  in low-mass supernovae as indicator. 
In the case of such a late input scenario, we can relate the measured abundance ratios to a single nucleosynthesis event. 
If this is the case, properties of this event can be inferred from measured abundance ratios. The main parameters are the delay time  $\Delta$ between the 
injection event an the condensation of the material into solids and the dilution factor $f$ that indicates to which extent the solar system material has been mixed with the ejecta from the last event.
Following \citet{Takigawa.Miki.ea:2008} and \citet{Banerjee.Qian.ea:2016} we can estimate the number ratio $N_R$/$N_S$ between a radioactive isotope with mass number $A_R$ and a stable reference nucleus with mass $A_S$ at solar system formation as
\begin{equation}
\label{eq:late_input}
\left(\frac{N_R}{N_S} \right)_{SSF} \approx \frac{f \times Y_R 
e^{-\Delta/\tau}
}{X_S^{\odot} M_\odot} \times\frac{A_R}{A_S}.
\end{equation} 
where $\Delta$ is the time between the last nucleosynthesis
event to produce the isotope of interest and the condensation of
the material into solid grains. $Y_R$ is the yield of the radioactive nucleus in
solar masses from our calculations and $X_S^{\odot}$ is the solar mass fraction of the
reference isotope, taken here from \citet{Lodders:2003}.
Due to its long half-life,  $^{92}$Nb is not very sensitive to $\Delta$ and therefore a good candidate
to constrain $f$.
Assuming
$\Delta\approx 1$~Myr and the lowest $^{92}$Nb yield from the
13~M$_\odot$ model we require a dilution factor
$f\approx 5 \times 10^{-3}$ to achieve the measured ratio of
$ 1.6 \times 10^{-5}$ \citep{Schoenbaechler.Rehkaemper.ea:2002}.
This value is significantly higher than what has been suggested 
in the literature \citep{Banerjee.Qian.ea:2016,Takigawa.Miki.ea:2008,Wasserburg.Busso.ea:2006}.
Furthermore, this value is also much larger than the upper limit of $f < 5 \times 10^{-4}$ that results 
from with the same model from the upper limit on the $^{98}$Tc/$^{96}$Ru ratio \citep{Becker.Walker:2003}.
Thus, in particular with the updated neutrino energies, it does not seem possible 
to explain the pre-solar abundance of  $^{92}$Nb with the input by a low mass supernova. 

\citet{Jacobsen.Matzel.ea:2009} have given a value of
$(17.2\pm2.5) \times 10^{-6}$ for the ratio $^{36}$Cl/$^{35}$Cl from
grains of the Allende meteorite and
\citet{Lin.Guan.ea:2005} have found a value of $5 \times 10^{-6}$
in material from the Ningqiang carbonaceous chondrite, giving a
combined range of possible values of roughly $(3-20)\times 10^{-6}$.
\cite{Jacobsen.Matzel.ea:2011} have suggested that 
late-stage irradiation of the proto-planetary disk is the most likely origin of
pre-solar  $^{36}$Cl while a stellar origin cannot be excluded.
Using again our 13~$M_\odot$ model with $\Delta=1$ Myr and $f= 5 \times 10^{-4}$ as suggested by 
\citet{Banerjee.Qian.ea:2016}
we find $^{36}$Cl/$^{35}$Cl ratios of $(8.3-9.5)\times 10^{-6}$ for low and high neutrino energies respectively compared to a ratio of   $6\times 10^{-6}$ for the yields without neutrinos.
All of these values are currently consistent with the range of observed ratios.
However,
the scaling of the $^{36}$Cl/$^{35}$Cl ratio with the neutrino energies make 
this nucleus an interesting candidate as neutrino thermometer if the parameters of
the late input scenario and the meteoritic ratio can be determined with better precision in the future.

\subsection{$\gamma$-ray sources $^{22}$\textup{Na} and $^{26}$\textup{Al}}
\label{sec:alna}
The characteristic $\gamma$-ray lines of the decay of long lived
${}^{26}\!$Al has allowed \citep{Diehl:2013} to estimate its present-day
equilibrium content in the Galaxy to be $2.8 \pm 0.8$ M$_\odot$. While
the sensitivity of ${}^{26}\!$Al yields from massive stars with respect to
thermonuclear reaction rates has been studied in detail by~\citet{Iliadis.Champagne.ea:2011}, we here explore the uncertainties
related to the $\nu$ process.

\begin{figure}[htb]
  \includegraphics[width=\linewidth]{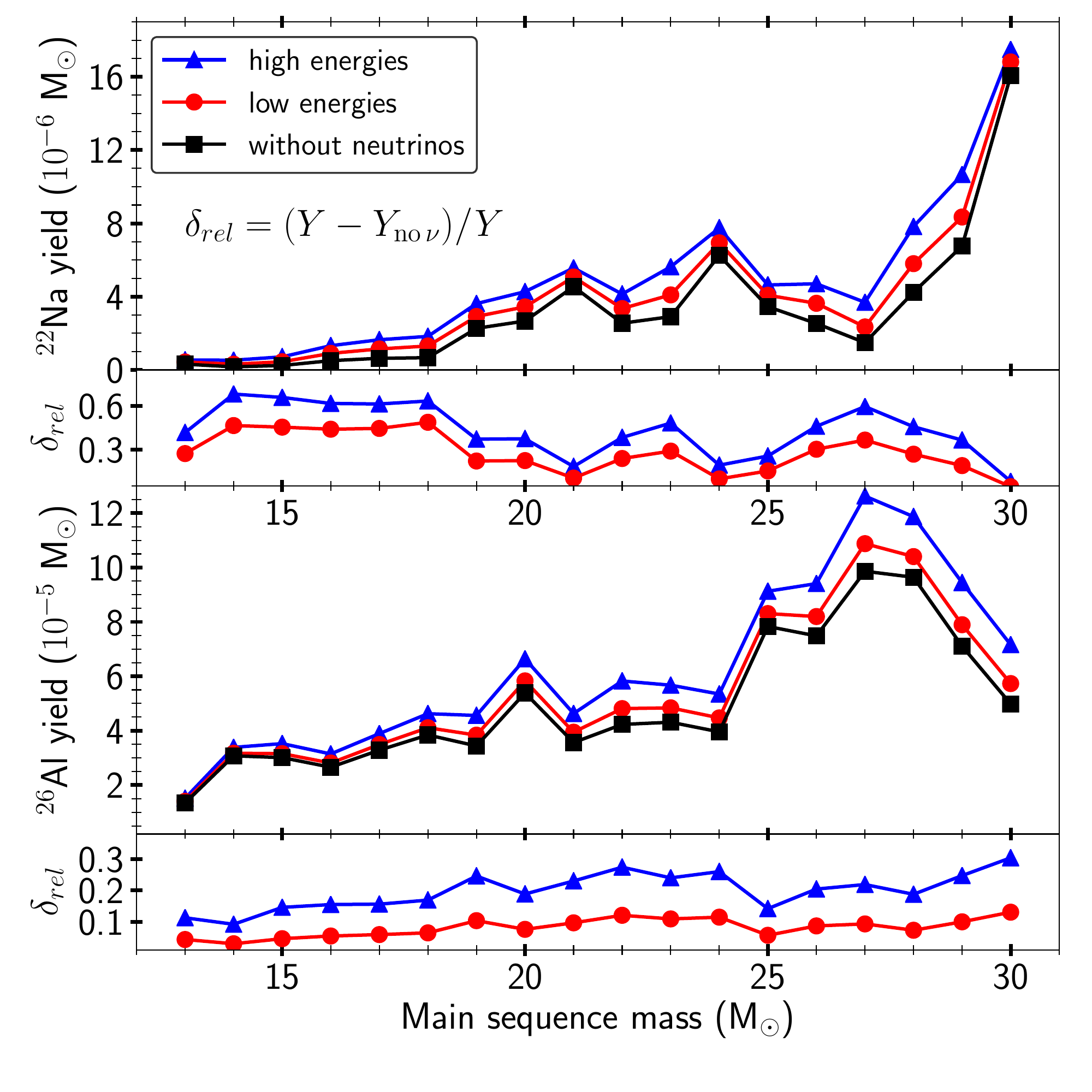}
  \caption{Yields of $^{22}$Na (upper panel) and ${}^{26}\!$Al (lower
   panel) for the set of progenitor stars considered here. Shown are
   the yields without neutrinos and including neutrinos with the
   high energies and the more realistic low energies.  Below
   each panel the relative differences
   $\delta_{rel}=(Y-Y_{\text{no}\nu})/Y$ between the yields with and
   without neutrinos are shown in smaller panels.  
\label{fig:al26na22}}
\end{figure}

The yield of ${}^{26}\!$Al is known to be enhanced by neutrino
nucleosynthesis~\citep{Woosley.Hartmann.ea:1990,Timmes.Woosley.ea:1995}.
We also find that the yield of ${}^{26}\!$Al is increased by factors
between 1.1 and 1.4 in the range of progenitor models studied (see
Table~\ref{tab:prodall} and Figure~\ref{fig:al26na22}). For low
energies the maximum increase is limited to a factor of 1.13 for the
most massive progenitor in our set. For the high neutrino energies, the enhancement is within
the precision of the galactic ${}^{26}\!$Al content estimate. 

\begin{figure*}[htb]
  \includegraphics[width=\linewidth]{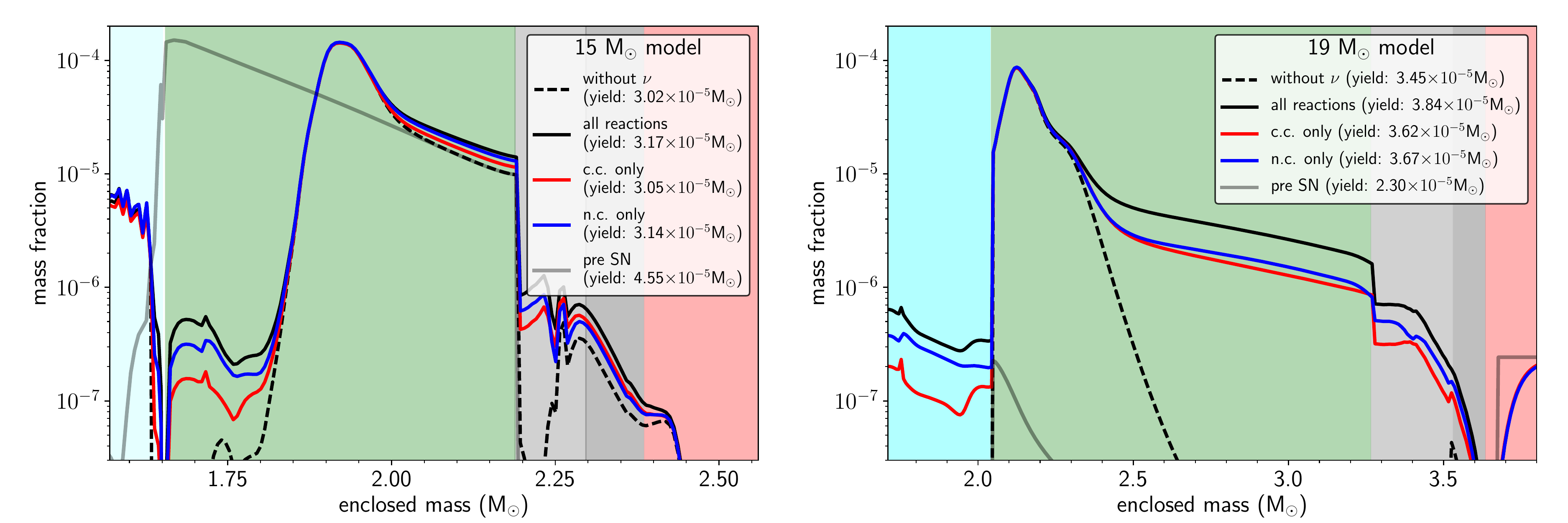}
  \caption{Mass fraction of ${}^{26}\!$Al for the 15 M$_\odot$ (left) and 19 M$_\odot$ (right) 
    progenitor models. Shown are the results for
    calculations with and without including neutrino interactions,
    with charged-current reactions only, and neutral-current reactions
    only for the low neutrino energies. The pre-supernova mass fraction are also
    shown. Comparison of the two models illustrates that the contribution from hydrostatic C-burning 
    almost irrelevant for the final yield. \label{fig:al26profiles}}
\end{figure*}

In massive stars ${}^{26}\!$Al can be produced in core H~burning,
C~burning in the core, and in a convective shell as well as during explosive
Ne/C~burning in the supernova shock. ${}^{26}\!$Al from H~core~burning survives largely in the envelope and is
partly blown away by stellar winds especially for more massive
stars. It contributes 25\%--55\% of the total yield with the exception
of the 13 M$_\odot$ progenitor for which the H~burning component
constitutes 70\%. This component is unaffected by the explosion and by
the neutrinos.

Figure \ref{fig:al26profiles} shows the ${}^{26}\!$Al mass fraction profile for the 15~M$_\odot$
model which contains a significant amount of ${}^{26}\!$Al from C-burning in the 
O/Ne shell before the supernova explosion. 
Most of this ${}^{26}\!$Al is destroyed by the shock, mainly by  
${}^{26}\!$Al$(n,p)^{26}$Mg and ${}^{26}\!$Al$(n,\alpha)^{23}$Na.
The 19 M$_\odot$ model shown in the right panel of Figure 
\ref{fig:al26profiles} differs significantly in its pre-supernova 
${}^{26}\!$Al mass fraction which is almost negligible.
However, both 
cases lead to a rather similar distribution and yield of  ${}^{26}\!$Al at the end.
When looking at the whole range of progenitors we find
that 
the final ${}^{26}\!$Al yield in the ejecta is effectively independent of the
inner C-burning component of pre supernova ${}^{26}\!$Al. Similar results
have been obtained by~\citet{Limongi.Chieffi:2006} for a different set
of progenitor models. 

Shock heating produces a peak in the mass fraction distribution during
explosive C/Ne burning. The reactions chain
$^{20}$Ne$(\alpha,\gamma)^{24}$Mg$(n,\gamma)^{25}$Mg$(p,\gamma){}^{26}\!$Al
competes with neutron induced reactions on ${}^{26}\!$Al and
photodissociation at higher temperatures. The optimal conditions for
the production of ${}^{26}\!$Al are found where the peak temperature
reaches around 2.3--2.5~GK, depending on the progenitors density
structure.  With the explosion model used here we find for all
progenitors a peak in the O/Ne shell.  Deeper inside, i.e.,  left of
the peak, no ${}^{26}\!$Al survives the shock heating while the $\nu$
process can operate further out.

Neutrinos contribute to the production of ${}^{26}\!$Al during the
explosive phase by two different mechanisms. Neutrino-induced
spallation reactions on the most abundant nuclei in the O/Ne shell,
$^{20}$Ne, $^{24}$Mg, and $^{16}$O increase the number of free
protons, enhancing the reaction $^{25}$Mg$(p,\gamma)$, which is also
the main production channel without neutrinos. Additionally, the
charged-current reaction $^{26}$Mg$(\nu_e,e^-)$ also contributes with
a cross section that is now based on experimental data as described in
\S\ref{sec:rates}. Figure~\ref{fig:al26profiles} illustrates the
neutral-current and charged-current contributions.  With the softer
neutrino spectra, we find that both charged- and neutral-current
reactions contribute to a similar extent to the production of
${}^{26}\!$Al in the O/Ne layer. The enhancement of the
$^{25}$Mg$(p,\gamma)$ is confined to a narrow region of optimal
temperature, whereas the $^{26}$Mg$(\nu_e,e^-)$ contributes more
evenly but to a lesser extent throughout the entire layer, decreasing
with the neutrino flux at larger radii.  The strength of the $\nu$
process also depends on the position of the ${}^{26}\!$Al production peak
within the O/Ne shell which in turn depends on the peak temperature.
The deeper inside the peak is, the more mass is on the right hand side
of the peak where the $\nu$~process can operate efficiently.  While
the total mass of the O/Ne layer scales with the initial progenitor
mass, the condition of the peak temperature is more sensitive to the
detailed structure of the individual stellar models.  Comparing the
two cases in Figure~\ref{fig:al26profiles} also illustrates this
dependence of the $\nu$~process region on the position of the
${}^{26}\!$Al peak.  The systematics of the total yield with respect to
the progenitor mass that are shown in Figure~\ref{fig:al26na22} follow
the trend of the $^{20}$Ne content of the pre-supernova models,
modulated by structural details affecting the position of the
${}^{26}\!$Al peak within the O/Ne layer.

The relative differences $\delta_{rel}$ due to neutrinos to Figure
\ref{fig:al26na22} can be understood from these aspects.  For the
13--18~M$_\odot$ progenitors, the ${}^{26}\!$Al peak is located in the
middle of the O/Ne shell, such that only a fraction of that shell is
cold enough for the $\nu$~process to contribute. Within this range of
progenitors the mass in the O/Ne shell increases giving rise to a
slight increase with progenitor mass. In all of these models a
substantial amount of ${}^{26}\!$Al is present from hydrostatic burning
but little of it survives the high temperatures during the explosion.  Starting from the 19 M$_\odot$ the
${}^{26}\!$Al peak is right at the bottom edge of the O/Ne shell. As
discussed by \citet{Woosley.Heger.Weaver:2002} energy generation in
central C~burning in stars heavier than this can no longer overcome
the neutrino losses which leads to substantial changes in structure
and nucleosynthesis, including a reduced abundance of ${}^{26}\!$Al in the
O/Ne shell.  In the mass range between 19--25~M$_\odot$ there is
basically no contribution from C burning and the $\nu$~process has the
strongest relative effect on ${}^{26}\!$Al because most of the O/Ne shell
is cold enough.  The 20 M$_\odot$ is a particular exception for which
a convective merger of shells has occurred and altered the structure
and composition \citep[see also][]{Woosley.Heger.Weaver:2002}.  The
relatively large yields for the 25--28 M$_\odot$ progenitor models
result from a drastic increase of the contribution from hydrostatic C
burning that decreases again in the 29 and 30 M$_\odot$ progenitors.
The 25--28 M$_\odot$ progenitors also exhibit the largest compactness
parameter~\citep{Oconnor.Ott:2011}
\begin{equation}
\xi_{2.5}=\frac{2.5
\text{M}_\odot}{R(M_r=2.5 \text{M}_\odot)/1000\text{km}}                                      
\end{equation}
in the range between 0.31-0.45 and also the largest pre-SN content of
$^{25}$Mg.  Note that according the explosion criterion suggested by
\cite{Ugliano.Janka.ea:2012}, stars with $\xi_{2.5}>0.35$ are likely
to fail to explode as supernovae.

The fact that the $\nu$~process mostly adds to ${}^{26}\!$Al in a
secondary way, i.e., by enhancing the abundance of protons, makes its
contribution to scale smoothly with the $\nu$ energy compared to the
weak dependence with the neutrino energy seen in
\S\ref{sec:lata} and \S\ref{sec:clnbtc}.

Since the position of the ${}^{26}\!$Al production peak depends on the peak temperature at that radius 
we can see from equation \ref{eq:Tpeak} that it also depends on the 
explosion energy. For less energetic explosions the optimal temperature is reached for smaller radii and a stronger impact of $\nu$~process can be expected 
because the neutrino fluxes are higher closer to the PNS.

The short-lived isomeric state of ${}^{26}\!$Al is not treated explicitly
in our calculation, but we assume a thermal equilibrium and use an
accordingly adjusted $\beta$-decay rate \citep{FFN1982}.  The
validity of this assumption in a supernova environment has been
confirmed by~\cite{Iliadis.Champagne.ea:2011}. 

Our results show that the major uncertainty for the yields of
${}^{26}\!$Al from massive stars originates from thermonuclear reaction
rates. \citet{Iliadis.Champagne.ea:2011} have estimated such a
uncertainty to be a factor 3. As experiment and theory advance, these
uncertainties are bound to shrink and the predictions can approach the
precision of the observations.

\begin{figure*}[tb]
  \includegraphics[width=\textwidth]{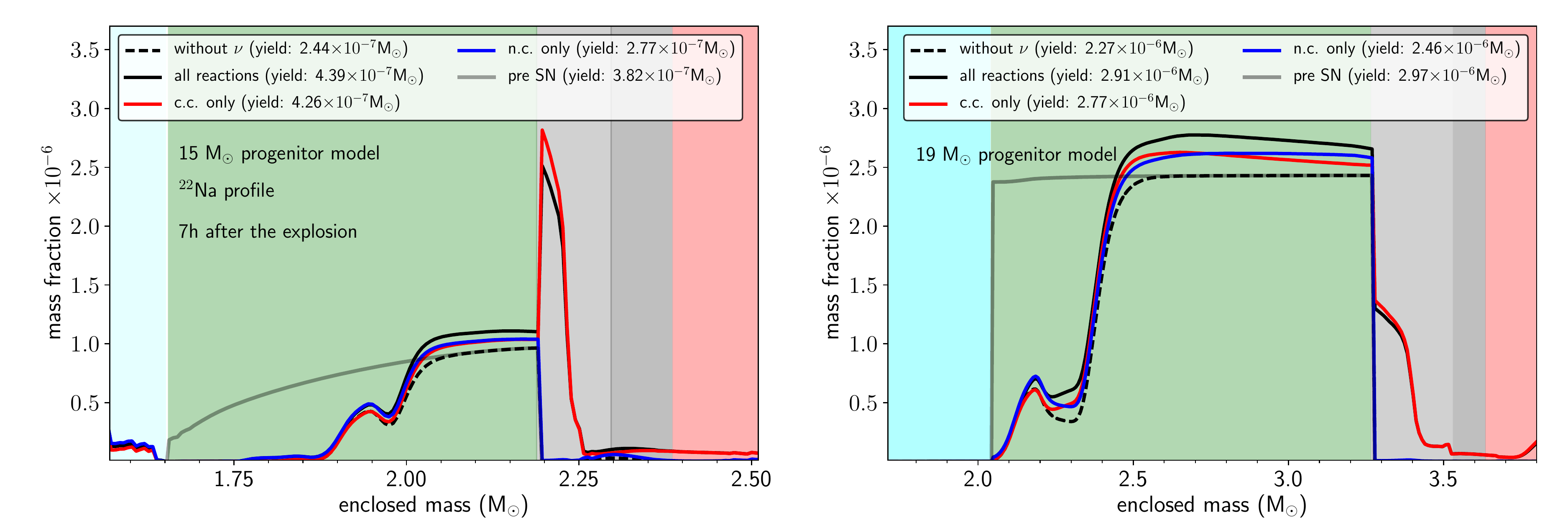}
  \caption{$^{22}$Na mass fraction profile for the 15 (left) and
    19~M$_\odot$ (right) progenitors for the set of low neutrino
    energies. While O/C shell contains the largest amount of $^{22}$Na
    the 15~M$_\odot$ model the contribution from this region is
    negligible for the 15~M$_\odot$
    progenitor. }\label{fig:na22profiles}
\end{figure*}

The radioisotope $^{22}$Na has a relatively short half-life of 2.6~yr
and decays to $^{22}$Ne emitting a positron followed by the emission of a $\gamma$ ray
line at $1.275$ MeV and two $0.511$~MeV $\gamma$ rays from the  annihilation of the positron. 
\citet{Woosley.Hartmann.ea:1989} have estimated the contribution
of  $^{22}$Na to the bolometric lightcurve and emission lines from
SN1987A based on a model that did not include the $\nu$~process. They conclude that a detection of the $\gamma$-ray line emission might become possible with future instruments.
In the following we describe how the $\nu$~process affects the production of $^{22}$Na for our range 
of progenitor models and discuss the detectability of this enhancement in photometry and as radiogenic  $^{22}$Ne in pre-solar grains. 

Figure \ref{fig:al26na22} shows that
supernovae could eject even larger amounts of $^{22}$Na but
representing a much smaller fraction of the total mass of the ejecta.
  The last
phases of shell burning produce mass fractions around
1--$4\times 10^{-6}$ of $^{22}$Na O/Ne shell. Without taking into
account the $\nu$ process the final ejected amount of $^{22}$Na is
only determined by how much of it survives the shock heating which
destroys $^{22}$Na for temperatures
above 1.8 GK. Just like in the case of ${}^{26}\!$Al the peak
temperatures reached in the O/Ne shell are the determining factor for
the ejected amount of $^{22}$Na. In general, there is no contribution
from pre-supernova wind and from the He shell or H envelope because
most of the $^{22}$Na that has been produced there during hydrostatic
burning has decayed at the time of core collapse.  Figure
\ref{fig:na22profiles} illustrates the distribution of the mass
fraction of $^{22}$Na for the 15 and 19 M$_\odot$ progenitor
models. At mass coordinates around 1.9 M$_\odot$ for the 15 M$_\odot$
model and around 2.1 M$_\odot$ for the 19 M$_\odot$ model there is
also a small peak in the mass fraction that results mostly from
$^{21}$Ne$(p,\gamma)^{22}$Na competing with the neutron captures.

The $\nu$ process affects this feature by changing the abundance of
free protons as in the case of ${}^{26}\!$Al discussed above.  This peak
is however always negligible compared to the bulk of the $^{22}$Na in
the outer part of the O/Ne shell which remains mostly unchanged by the
passage of the supernova shock without taking into account neutrinos.
The $\nu$ process liberates free protons that increase of the
$^{22}$Na mass fraction in the outer O/Ne shell as can be seen in
Figure \ref{fig:na22profiles} and also the charged current reaction
$^{22}$Ne$(\nu_e,e^-)^{22}$Na contributes.
$^{23}$Na$(\nu,\nu' n)^{22}$Na has been suggested by
\cite{Woosley.Hartmann.ea:1990} as an additional source of ${22}$Na. We
find that this channel contributes 50\% of the neutral current effects
for the 16~M$_\odot$ model.  Figure \ref{fig:na22profiles}
illustrates that both channels contribute to about the same extend in
the outer O/Ne shell. This can be understood from the pre-supernova
composition because the O/Ne shell consists of roughly equal mass
fractions of $^{22}$Ne and $^{21}$Ne that range between
$5 \times 10^{-5}$ to $5\times10^{-4}$ in the O/Ne shell of this progenitor.

More striking is the effect of the charged current
$^{22}$Ne$(\nu_e,e^-)^{22}$Na reaction that increases the mass fraction of
$^{22}$Na in the O/C shell and produced the very prominent peak for the
15~M$_\odot$ progenitor that can be seen in the left panel of Figure
\ref{fig:na22profiles}.  In contrast, this production channel is negligible in
the 19~M$_\odot$ model shown in the right panel.  The O/C shell contains very
little $^{21}$Ne and therefore thermonuclear and also neutral current
contributions to $^{22}$Na are suppressed.  The mass fraction of $^{22}$Ne in
the O/C shell which has not undergone C-burning is between $1-1.5 \times
10^{-2}$ and very similar for all progenitor models studied here.  Still, only
the 14--18~M$_\odot$ models show a major production of $^{22}$Na in the O/C
shell due to the $\nu_e$ capture on $^{22}$Ne that contributes at least 80\% of
the total $\nu$ process enhancement for the progenitors in that range.  That is
reflected in larger values for $\delta_{rel}$ in Figure \ref{fig:al26na22}.  As
described above,
the energy balance of C burning changes and consequently also the
star's structure, when going from the 18 to the 19 M$_\odot$ model.
While the inner edge of the O/C shell is located between 17,000 and
20,000~km for progenitor models below 19 M$_\odot$, its position moves
out to more than 30,000 km for most of the more massive models. This
reduces the maximum neutrino flux by more than a factor two and the
neutrino induced production is suppressed.  The abundance of $^{22}$Na
in the outer part of the O/Ne shell is then much larger than at the
bottom of the O/C shell for the more massive models. The 27 M$_\odot$
model is again an exception for which the abundances in both regions
are similar again. That is because the 27 M$_\odot$ model has very
little $^{4}$He left in the O/C shell, such that the neutron
production via $^{22}$Ne$(\alpha,n)$, which drives the destruction of
$^{22}$Na by neutron captures, is suppressed. For all the cases studied here, the  $^{22}$Na yield with only charged current
reactions is larger than with only neutral current reactions.  For the
low neutrino energies the charged current alone contributes at least 70\% of
the total $\nu$~process enhancement and for the higher energies it is at
least 60\%.

Assuming a total yield of $2\times 10^{-6}$ M$_\odot$ of  $^{22}$Na \citet{Woosley.Hartmann.ea:1990} found that the contribution of the $^{22}$Na decay to the supernova lightcurve is of the order of $10^{36}$ erg/s, very similar to the contribution from  $^{44}$Ti decay during the first 2-3 years. During this time, the decay of  $^{56}$Co still dominates the bolometric luminosity with $10^{40}$-$10^{38}$ erg/s. Later $^{44}$Ti with a half-life of 59~years dominates the luminosity while most of the $^{22}$Na has already decayed.
 Therefore, we do not expect the enhancement of  $^{22}$Na due to the $\nu$~process to make a difference for the bolometric lightcurve of a supernova (see also \citet{Kozma.Fransson:1998} and \citet{Diehl.Timmes:1998}). 
 The $\gamma$-ray line at $1.275$ MeV has been detected with the COMPTEL experiment
on board the Compton Gamma-Ray Observatory associated with Nova
Cassiopeia 1995. From this observation \citet{Iyudin:2010} has
estimated the total amount of ejected $^{22}$Na to be
$\approx 10^{-7}$ M$_\odot$ with large uncertainties remaining due to
the distance ($\approx 3$kpc) and total ejected mass
($\approx 10^{-3}$ M$_\odot$).  
 Scaling the results from \citet{Woosley.Hartmann.ea:1989} for a supernova at a distance of 10 kpc with an escape fraction of 20\% at 400 days after the explosion we find a photon flux of 
 $5.3\times 10^-6$ cm$^{-2}$s${-1}$ per $10^{-6}$ M$_\odot$ of 
  ejected $^{22}$Na. 
   \citet{Teegarden.Watanabe:2006} give the sensitivity of the SPI $gamma$-ray telescope on the INTEGRAL satellite as $1.2\times 10^{-4}$ for the 1.275 MeV $\gamma$-ray line. 
   The expected photon flux for the largest amount of  $^{22}$Na we find for the 30 M$_\odot$ model would barely lead to such a photon flux. 
   In order to distinguish between the 
  low and high energy scenario discussed here, we would require to resolve a flux difference of $5 \times 10^{-6}$ photons cm$^{-2}$s${-1}$ which might become feasible with future 
  space based $\gamma$-ray telescopes like the proposed e-ASTROGAM mission \citep{e-astrogam}.
  
In addition to the emission in the electromagnetic spectrum $^{22}$Na might also be relevant as
the progenitor of $^{22}$Ne found in meteoritic grains. 
\citet{Clayton:1975} has already pointed out that the $^{22}$Ne-rich
Ne-E(L) component in low density graphite grains from meteorites first
found by \citet{Black.Pepin:1969} could be a consequence of $^{22}$Na
decay, i.e., the Ne found in these grains would be pure $^{22}$Ne
originally condensed as $^{22}$Na. More recently \citet{Amari:2009}
has concluded that the O/Ne shell of massive stars are the most likely
origin of the material with very low $^{20}$Ne/$^{22}$Ne ratios below
$0.01$. The condensation of graphite grains in O-rich material is
problematic \citep{Lattimer.Schramm.ea:1978} even though models exists
that would allow for it \citep{Clayton.Liu.ea:1999}.  The C/O ratio in
the O/Ne shell is typically $\mathrm{C/O}\approx 0.01$ while this
ratio reaches $\mathrm{C/O} \approx 0.3$ in the O/C shell where
charged current reactions produce most of the $^{22}$Na.  Modest
mixing with the C/O shell right on top of it could easily lead to
material satisfying $\mathrm{C/O}>1$ and strongly enriched in
$^{22}$Na.  The $\nu$~process allows for the production of a large
fraction of $^{22}$Na in more C-rich supernova ejecta but the ratio of
$^{12}\textrm{C}/^{13}\textrm{C}\approx 10^4$--$10^5$ in these layers
still requires mixing with the outer He or H rich layers to explain
the high $^{12}$C/$^{13}$C ratio of $313\pm2$ found in the same grains
\citep{Meier.Heck.ea:2012}.

$^{44}$Ti has been detected in supernova
remnants~\citep{Iyudin.Diehl.ea:1994,Grefenstette.Harrison.ea:2014}. It
is produced mainly in the inner ejecta in an $\alpha$-rich freeze out
of NSE~\cite{Woosley.Heger.Weaver:2002}. At high temperatures, photon-
and charged particle induced reactions dominate over any neutrino
contribution. Therefore, we find no significant effect of neutrinos on
the yield of $^{44}$Ti.  The production of $^{60}$Fe in supernovae is
discussed in detail by~\cite{Limongi.Chieffi:2006}, where
the neutron density reached during the shock is identified as a key
parameter for the yield. Despite the increase in the density of free
nucleons due to neutrino spallation reactions, we find no significant
modification of the $^{60}$Fe yield because neutrons are mostly
captured by other, in particular heavier nuclei.

\section{Conclusions}
\label{sec:conclusions}
We have performed an updated study of $\nu$ process nucleosynthesis
taking into account for the first time the results from recent
supernova
simulations~\citep{Huedepohl.Mueller.ea:2010,Martinez-Pinedo.Fischer.ea:2012,Martinez-Pinedo.Fischer.Huther:2014}
that predict noticeably lower average energies particularly for $\mu$
and $\tau$ (anti)neutrinos. As a results we found charged current
processes to be now more relevant.
Compared to previous studies, we use a full set of neutrino-induced
charged- and neutral current reactions including spallation products
for all nuclei in our reaction network with charge numbers
$Z<76$. Where cross sections derived from experimental data or dedicated shell model calculations are available we use those and we have included additional experimental data to infer the low energy part of cross sections for several charged current processes. This extensive compilation of cross sections for neutrino induced reactions will be published along with this article
and can then be employed for the calculation of the next generation of stellar yield tables.\\
We have performed this study for a range of progenitor models for
massive stars with ZAMS masses in the range between 13 and 30
M$_\odot$, highlighting sensitivities and trends with respect to
stellar structure and composition.  Our nucleosynthesis study confirms
the contribution of the $\nu$ process to the production of $^{11}$B,
$^{138}$La, and $^{180}$Ta and avoids the overproduction of these
elements that has been found in previous studies.
Furthermore, we discuss the interplay between
$\gamma$~process and $\nu$~process production and find that for
individual progenitor models, in particular for the 27~M$_\odot$
model, neutral~current neutrino interactions leading to the emission
of neutrons have a major effect on the production of $^{180}$Ta for some progenitor models.

We confirm that the $\nu$~process cannot be the primary origin of
$^{19}$F and emphasize remaining uncertainties with
respect to thermonuclear reaction cross-sections and stellar structure.
Moreover, we find that there is no nucleus for which the $\nu$ process
can be assumed as the only origin. Unless all other contributions to the solar inventory are well understood it
is therefore near impossible to use comparisons to the solar abundances to give stringent constraints on neutrino properties.
If we consider the scenario of a nearby supernova explosion polluting the pre-solar cloud with short-lived
radioactive nuclei, we find that the $^{36}$Cl/$^{35}$Cl ratio that 
can be measured in meteoritic grains is sensitive to the $\bar{\nu_e}$ neutrino spectra. 
In this case, the $^{36}$Cl/$^{35}$Cl ratio could be used as a ``neutrino thermometer''.

Including neutrino reactions with on all nuclei in our network we have
identified effects on nuclei on the $10$\% level that have not been discussed
before in the literature, including the p-nucleus  $^{113}$In.  We also find
modifications of the yields of $^{33}$S,$^{40}$Ar,$^{41}$K and Fe-group nuclei
that originate from the Si/O shell. Quantitatively, those results need to be
taken with caution due to the limits of our 1D supernova model. We also have included reactions
suggested by \citet{Heger.Kolbe.ea:2005} and find their effects to be small. 
We conclude
that for the range of supernova models the most important effects of the
$\nu$~process on the production of stable nuclei have been identified.  In the
second part we discuss how neutrino-induced reactions, directly and indirectly,
contribute to the production of long- and short-lived radioactive nuclei.  For
$^{92}$Nb and $^{98}$Tc we also discuss the competition thermonuclear and
neutrino induced production.  Within our model we cannot explain the
$^{92}$Nb/$^{93}$Nb ratio found in meteoritic grains but due to the sensitivity
of the yields to nuclear reaction rates and stellar structure and potential
contribution from neutrino driven winds further studies are required.

The yields of $^{22}$Na and ${}^{26}\!$Al, both prime candidates for
gamma-ray astronomy, are enhanced.  For ${}^{26}\!$Al the magnitude of
this enhancement is of the order of a few \%. 
We also find that
significantly larger uncertainties due to nuclear reaction
rates remain.
Even though the enhancement of the production of $^{22}$Na is of the order of 50\% we do not expect 
a direct effect on the supernova lightcurve and we estimate that the fluxes of characteristic $\gamma$ ray emission are
below the limit of the sensitivity of current instruments. However, the production of $^{22}$Na in the O/C leads us to  suggest the $\nu$~process as the
origin of the Ne-E(L) component found in low density graphite grains.

Many of the relevant neutrino-nucleus cross-sections rely almost
entirely on theoretical calculations and are therefore accompanied by
large uncertainties. Experimental data on the relevant transitions
could help to reduce the uncertainties in order to make inferences
from observations more reliable. Furthermore, important uncertainties
remain related to the progenitor
structure~\citep{Woosley.Heger.Weaver:2002}, helium burning
rates~\citep{Austin.West.Heger:2014}, and the long term evolution of
the neutrino spectra and neutrino oscillations~\citep{Wu.Qian.ea:2015}.

\begin{acknowledgments}
  We thank Yong-Zhong Qian, Projjwal Banerjee and Meng-Ru Wu for
  useful discussions.  This work was partly supported by the Deutsche
  Forschungsgemeinschaft through contract SFB~1245 and the Helmholtz
  Association through the Nuclear Astrophysics Virtual Institute
  (VH-VI-417). AH was supported by an Australian Research Council
  (ARC) Future Fellowship (FT120100363) and by the US National Science
  Foundation under Grant No. PHY-1430152 (JINA Center for the
  Evolution of the Elements).
\end{acknowledgments}


\end{document}